\documentclass[a4paper]{jpconf}
\usepackage{graphicx}
\usepackage{refstyle}
\usepackage{amssymb}
\usepackage{amsmath}
\usepackage{bm}
\usepackage{braket}
\usepackage[latin9]{inputenc}

\makeatletter

\AtBeginDocument{}
\RS@ifundefined{subref}
  {\def\RSsubtxt{section~}\newref{sub}{name = \RSsubtxt}}
  {}
\RS@ifundefined{thmref}
  {\def\RSthmtxt{theorem~}\newref{thm}{name = \RSthmtxt}}
  {}
\RS@ifundefined{lemref}
  {\def\RSlemtxt{lemma~}\newref{lem}{name = \RSlemtxt}}
  {}

\makeatother

\global\long\def\braOket#1#2#3{\left\langle #1\middle|#2\middle|#3\right\rangle }
\global\long\def\ini{\Psi_{{\rm ini}}}
\global\long\def\fin#1{\Psi_{c|#1}}
\global\long\def\ket#1{\left|#1\right\rangle }

\begin{document}

\title{Different interpretations of quantum mechanics make different predictions in non-linear quantum mechanics, and some do not violate the no-signaling condition}

\author{Bassam Helou and Yanbei Chen}
\address{Institute of Quantum Information and Matter, and Burke Institute of Theoretical Physics, M/C 350-17, California Institute of Technology, Pasadena, CA 91125, USA}
\ead{bhelou@caltech.edu}

\begin{abstract}
Nonlinear modifications of quantum mechanics have a troubled history. They were initially studied for many promising reasons: resolving the measurement problem, testing the limits of standard quantum mechanics, and reconciling it with gravity. Two results substantially undermined the credibility of non-linear theories. Some have been experimentally refuted, and more importantly, all deterministic non-linear theories can be used for superluminal communication. 
However, these results are unconvincing because they overlook the fact that the distribution of measurement results predicted by non-linear quantum mechanics depends on the interpretation of quantum mechanics that one uses. For instance, although the Everett and Copenhagen interpretations agree on the expression of Born's rule for the outcomes of multiple measurements in linear quantum mechanics, they disagree in non-linear quantum mechanics. 
We present the range of expressions of Born's rule that can be obtained by applying different formulations of quantum mechanics to a class of non-linear quantum theories.
We then determine that many do not allow for superluminal communication but only two seem to have a reasonable justification.
The first is the Everett interpretation, and the second, which we name \emph{causal-conditional}, states that a measurement broadcasts its outcome to degrees of freedom in its future light-cone, who update the wavefunction that their non-linear Hamiltonian depends on according to this new information.
\end{abstract}

\section{Introduction }

Non-linear quantum mechanics (NLQM) has long been considered as a possible generalization of standard quantum mechanics (sQM)  \cite{BirulaNLQM,debroglie,WEINBERGNLQM}, for three main reasons. First, the measurement process is controversial. If we assume that linear quantum mechanics explains all processes, then it is very difficult to explain wavefunction collapse \cite{BassiMeasurementProblem}. Phenomenological non-linear, stochastic, and experimentally falsifiable extensions of quantum mechanics (QM) have been proposed to explain the measurement process \cite{BassiCollapseReview},
and upper bounds on the parameters of such theories have been obtained
in \cite{BassiCollapseReview, helouSN,nanoCantileverTest}. Second, we would like to test the domain of validity of sQM. One possible feature to test
is linearity. Experimental tests of certain non-linear theories
have been performed in  \cite{neutronNLsearch,WinelandNLtest,ChuppNLtest,WalsworthTest,MajumderTest},
and all have returned negative results. Third, non-linear and deterministic theories
of QM have been proposed to combine quantum mechanics
with gravity. For instance, the Schroedinger-Newton theory describes
a classical spacetime which is sourced by quantum matter \cite{huanSN},
and the correlated worldlines theory is a quantum theory of
gravity which postulates that gravity correlates quantum trajectories in the path integral \cite{CWL}.

NLQM became a much less credible theory after 1990 because Gisin showed that deterministic NLQM could allow for superluminal communication \cite{GisinNLQMsuperLuminal}. The
no-signaling condition states that one cannot send information faster than the speed of light, and is a cornerstone of the special theory of relativity. The community regards the condition as being inviolable.
Gisin's work was quickly followed by others with similar conclusions  \cite{polchinskyEPR,Czachor1991}. Additional work then showed that under general conditions NLQM allows for superluminal communication \cite{SimonNoSignaling,Mielnik20011}.

In \cite{helouSN}, we showed that NLQM suffers from another serious conceptual issue: Born's rule cannot be uniquely extended from sQM to NLQM. 
Born's rule provides a prescription for predicting the distribution of measurement results in a particular experiment, and has, so far, passed all experimental tests. Any non-linear theory must make predictions that become equivalent to Born's rule in sQM when the non-linearity vanishes. 

%As in sQM, time-evolving the wavefunction in NLQM is a well-defined operation that is performed via a well-defined partial differential equation. 
As in sQM, measurements in NLQM pose significant conceptual difficulties. Fortunately, in sQM, these difficulties do not result in practical challenges. Whether the experimentalist is an Everettian or a proponent of wavefunction collapse does not matter, as in both cases they can safely use Born's rule to  predict the distribution of measurement results. This is no longer true in NLQM. Different interpretations of quantum mechanics result in different predictions for the outcome of an experiment, and so result in different expressions for Born's rule.

In this article, we search for interpretations of quantum mechanics that do not violate the no-signaling condition when are applied to NLQM. Since there could be interpretations that haven't been discovered, our approach won't be to extend all known interpretations to NLQM. Instead, we will extend the mathematical expression of Born's rule in a general way to NLQM, without regards to interpretation. After finding causal prescriptions, we speculate about their interpretation. 

Note that our analysis doesn't cover all possible non-linear theories. We only wish to show that non-linear quantum mechanics does not necessarily violate the no-signaling condition. More importantly, how to write down a general non-linear modification of quantum mechanics is still an open question. For example, the class of non-linear theories proposed by Weinberg in \cite{WEINBERGNLQM} doesn't include P.C.E. Stamp's proposal in \cite{CWL}. 
%In the language of Weinberg's NLQM paper, we limit ourselves to bilinear functions.
We also do not, \textit{a priori}, place any physical constraints on the class of non-linear theories we study. 
%as perhaps this is controversial and will make our work even more complicated. The interested reader can place them themselves
We only place one mathematical constraint: a single Dirichlet boundary condition is enough to completely specify a solution.
%\footnote{Although we believe that our work can be extended to theories where a finite number of boundary conditions is needed to specify a solution.}.

This article is outlined as follows. By introducing the formalism for multiple measurements in sQM, we show that linearity prevents two parties from communicating with each other faster than the speed of light. We then motivate the dependence of NLQM on the formulation of quantum mechanics by providing a simple example involving a single measurement.
By extending the notion of a time-evolution operator to NLQM, we generate extensions of Born's rule in the context of multiple measurements.  
Afterwards, we discuss what well-known formulations of quantum mechanics, such as the Everett interpretation, predict for the distribution of measurement results in NLQM.
We then present all possible prescriptions that do not violate the no-signaling condition. Finally, we propose and discuss a sensible prescription, which we name \emph{causal-conditional}, that doesn't violate the no-signaling condition. It states that a measurement broadcasts its outcome to degrees of freedom in its future light-cone, who update the wavefunction that their non-linear Hamiltonian depends on according to this new information.
%for assigning boundary states to time-evolution operator that doesn't violate the no-signaling condition.

\section{Multiple measurements in sQM and the no-signaling condition}

\begin{figure}
\centering\includegraphics[scale=0.4]{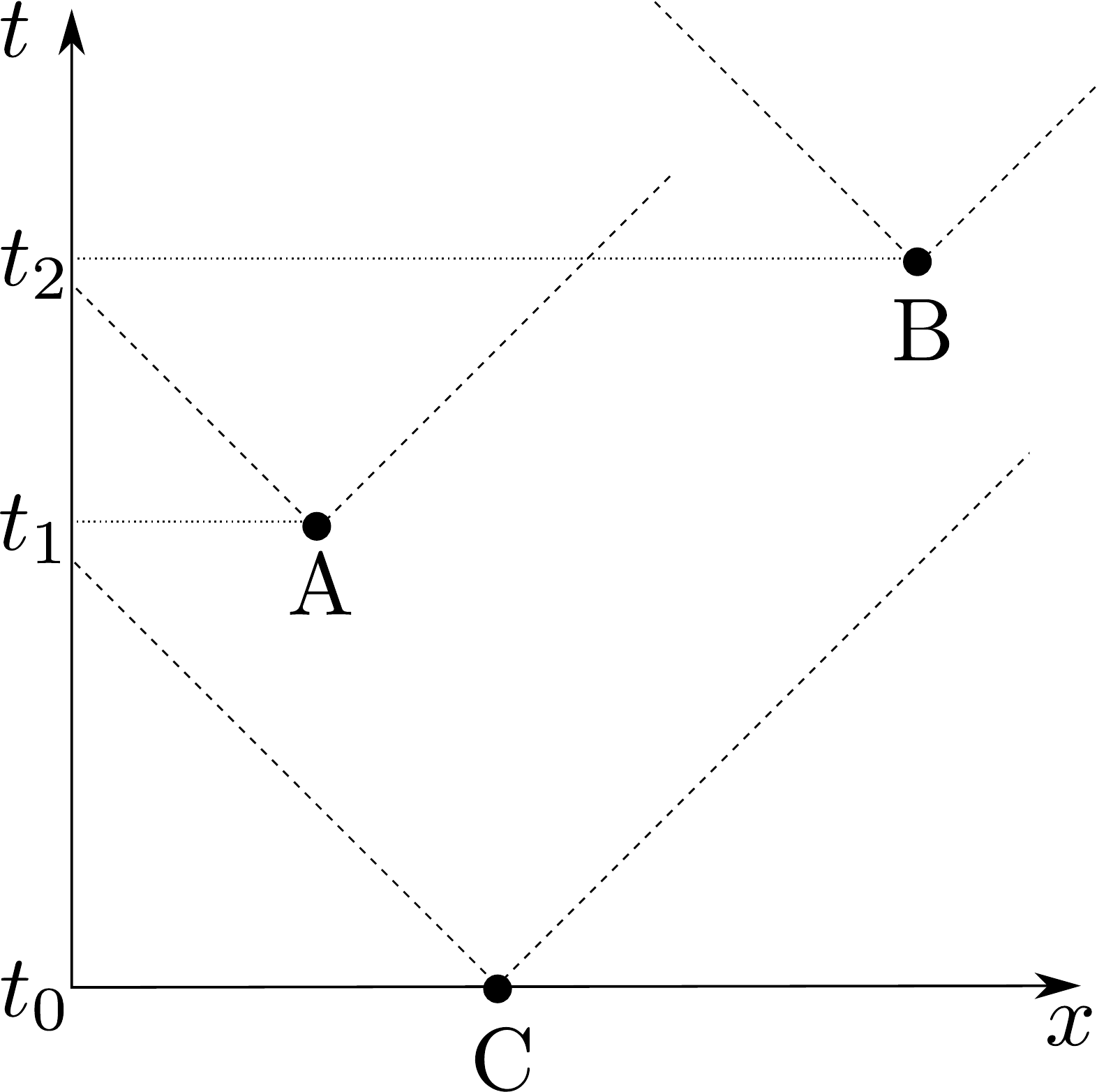}

\caption{\label{fig:setup_measurements}A spacetime diagram showing multiple measurement
events. Event C describes the preparation
of an ensemble of identical 2-particle states $\left|\Psi_{{\rm ini}}\right\rangle $
by Charlie. Event A (B) describes Alice (Bob) measuring her (his)
particles. The dashed lines show the light cone centered around each
event. }

\end{figure}

In Fig. \ref{fig:setup_measurements}, we show the setup that is typically used to show that NLQM violates the no-signaling condition. Charlie prepares a collection of identical arbitrary 2-particle states $\left|\ini\right\rangle $, and then sends them to Alice and Bob, such that they each hold one part of each of the states $\left|\Psi_{{\rm ini}}\right\rangle $. Alice performs measurements on her ensemble of particles at time $t_{1}$, and then Bob on his at a later time $t_{2}$.  We assume that Alice's measurements are space-like separated from Bob's, and so their particles do not interact from $t_1$ till $t_2$.

\subsection{Born's rule for multiple measurements}

Denote the probability that Alice measures $\alpha$ in some basis $A_a$, and Bob measures $\beta$ in some basis $B_b$ by $p\left(\alpha,\beta\right|A_a,B_b)$. For example, if the particles were
spins, $A_a$ could be the $\hat{\sigma}_{z}$ eigenstates, $\left|\uparrow\right\rangle ,\left|\downarrow\right\rangle $, and $B_b$ the $\hat{\sigma}_{x}$ eigenstates
$\left|\pm\right\rangle =\left(\left|\uparrow\right\rangle \pm\left|\downarrow\right\rangle \right)/\sqrt{2}.$
We will first determine $p\left(\alpha,\beta\right|A_a,B_b)$ according to sQM, and then discuss the different ways of generalizing it to NLQM in the next section.

In sQM, $p\left(\alpha,\beta\right|A_a,B_b)$ is given by 
\begin{equation}
p\left(\alpha,\beta\right|A_a,B_b) = \left\langle \fin{\alpha,\beta}|\fin{\alpha,\beta}\right\rangle ,\label{eq:expressionProb}
\end{equation}
where $\left|\fin{\alpha,\beta}\right\rangle $ is the unnormalized
joint quantum state of Alice and Bob at $t_{2}$, conditioned on the measurement results
$\alpha$ and $\beta$: 
\begin{equation}
\left|\Psi_{c|\alpha,\beta}\right\rangle =\left(\hat{I}_A\otimes\hat{P}_{\beta}\right)\hat{U}\left(t_{2},t_{1}\right)\left(\hat{P}_{\alpha}\otimes\hat{I}_B\right)\hat{U}\left(t_{1},t_{0}\right)\left|\Psi_{{\rm ini}}\right\rangle,
\label{eq:condStateComplicated}
\end{equation}
where $\hat{I}_A$ ($\hat{I}_B$) is the identity operator acting on Alice's (Bob's) particle, $\hat{U}(t,z)$ is the total time evolution operator for both
Alice and Bob's particles from times $z$ till $t$. The projectors are $\hat{P}_{\alpha} = \left|\alpha\right\rangle \left\langle \alpha\right|$ and $\hat{P}_{\beta} = \left|\beta\right\rangle \left\langle \beta\right|$.
To simplify the formalism, we only work with pure states. Our analysis is general because $\left|\Psi_{{\rm ini}}\right\rangle $ can always be enlarged to include the initial state of the environment.

Alice and Bob's measurement events are spacelike separated, so from $t_1$ till $t_2$, the interaction Hamiltonian between Alice's and Bob's particle is zero, and $\hat{U}(t_2,t_1)$ is separable
\begin{equation}
\hat{U}(t_2,t_1) \equiv \hat{A}(t_2,t_1) \hat{B}(t_2,t_1). 
\end{equation}
$\hat{A}$ acts on Alice's particle's Hilbert space, and $\hat{B}$ on Bob's. We can then rewrite Eq. (\ref{eq:condStateComplicated}) to 
\begin{equation}
\left|\Psi_{c|\alpha,\beta}\right\rangle =\left(\hat{I}\otimes\hat{P}_{\beta}\right)\hat{A}(t_2,t_1)\hat{B}(t_2,t_1)\left(\hat{P}_{\alpha}\otimes\hat{I}\right)\hat{U}\left(t_{1},t_{0}\right)\left|\Psi_{{\rm ini}}\right\rangle,
\label{eq:condState}
\end{equation}
which we substitute into Eq.  (\ref{eq:expressionProb})
\begin{equation}
p\left(\alpha,\beta\right|A_a,B_b)=\braOket{\ini}{\hat{U}^{\dagger}(t_1,t_0)\hat{P}_{\alpha}\hat{B}^{\dagger}(t_2,t_1)\hat{P}_{\beta}\hat{B}(t_2,t_1)\hat{P}_{\alpha}\hat{U}(t_1,t_0)}{\ini},
\label{eq:pABlinear}
\end{equation}
where we've used that $\hat{A}(t_2,t_1)$ commutes with $\hat{P}_{\beta}$.
Since Alice and Bob's measurement events are spacelike separated, $\hat{P}_{\alpha}$ and $\hat{P}_{\beta}$ commute,
\begin{equation}
\left[\hat{P}_{\alpha},\hat{P}_{\beta}\right]=0,
\label{eq:PaPbcomm}
\end{equation}
which we use to simplify Eq. (\ref{eq:pABlinear}) to 
\begin{equation}
p\left(\alpha,\beta\right|A_a,B_b)=\braOket{\ini}{\hat{U}^{\dagger}(t_1,t_0)\hat{B}^{\dagger}(t_2,t_1)\hat{P}_{\beta}\hat{B}(t_2,t_1)\hat{P}_{\alpha}\hat{U}(t_1,t_0)}{\ini}.
\label{eq:pABLinSimple}
\end{equation}
%Note that the above expression only holds for reference frames where Alice's measurement event occurs before Bob's. In other reference frames, we have to reverse their roles.

\subsection{\label{sec:No-signaling-condition-for-general-theories}The no-signaling
condition}

Superluminal communication from Alice to Bob is possible when
\begin{equation}
p\left(\beta\right|A_a, B_b)=\sum_{\alpha}p\left(\alpha,\beta\right|A_a,B_b)
\end{equation}
is influenced by Alice's choice of a measurement basis in a deterministic way. Since Bob can easily estimate $p\left(\beta\right|A_a, B_b)$, he can determine the basis Alice chose for her measurement results, which can form the foundation of a communication strategy. 
For instance, both Alice and Bob can agree that a particular choice of Alice's measurement basis could be associated with sending the bit 0, while another choice could be associated with the bit 1.

In sQM, superluminal communication can never occur because, using Eq. (\ref{eq:pABLinSimple}),
\begin{eqnarray}
p\left(\beta\right|A_a, B_b) & = & \braOket{\ini}{\hat{U}^{\dagger}(t_1,t_0)\hat{B}^{\dagger}(t_2,t_1)\hat{P}_{\beta}\hat{B}(t_2,t_1)\left(\sum_{\alpha}\hat{P}_{\alpha}\right)\hat{U}(t_1,t_0)}{\ini} \\
& = & \braOket{\ini} {\hat{U}^{\dagger}(t_1,t_0)\hat{B}^{\dagger}(t_2,t_1)\hat{P}_{\beta}\hat{B}(t_2,t_1)\hat{U}(t_1,t_0)}{\ini},
\end{eqnarray}
is clearly independent of $A_a$. We've shown that $p\left(\beta\right|A_a, B_b) = p\left(\beta\right|B_b)$, so sQM doesn't violate the no-signaling condition.

\section{Ambiguity of Born's rule in NLQM}

In this section, we discuss the ambiguity of Born's rule in NLQM. We first present an  example with one measurement, and then present a general formalism for generating prescriptions for calculating the distribution of the outcomes of an arbitrary number of measurements at arbitrary spacetime points.

\subsection{A simple example}

For experiments with a single measurement, sQM states that the probability of measuring an outcome $f$ is
\begin{equation}
p_{i\rightarrow f}=\left|\braOket f{\hat{U}}i\right|^{2},\label{eq:preSimple}
\end{equation}
where $\ket{f}$ is the pointer state associated with the outcome $f$.
The expression (\ref{eq:preSimple}) is usually interpreted in the following way. A preparation device prepares the system in $\ket{i}$, which evolves for some period of time under the time-evolution operator $\hat{U}$. The system then interacts with the measurement device. What happens next depends on one's interpretation of quantum-mechanics. An Everettian would state that decoherence splits the wavefunction into numerous branches that are approximately classical. On the other hand, a proponent of objective collapse would state that, due to stochastic modifications of the Schroedinger equation that become important when a microscopic system interacts with a macroscopic one, the wavefunction collapses onto $\ket{f}$ with probability $p_{i\rightarrow f}$. As in Ref. \cite{helouSN}, we will refer to the formulation of Born's rule based on $p_{i\rightarrow f}$ as \emph{pre-selection}.

Eq. (\ref{eq:preSimple}) admits even more interpretations. For instance, it can be rewritten as
\begin{equation}
p_{i\leftarrow f}=\left|\braOket i{\hat{U}^{\dagger}}f\right|^{2},\label{eq:postSimple}
\end{equation}
which can be interpreted as $\ket{f}$ evolves backwards in time to $\hat{U}^{\dagger}\ket{f}$ and is then projected by the preparation device to the state $\ket{i}$. We will refer to the formulation of Born's rule based on $p_{i\leftarrow f}$ as \emph{post-selection}.

In NLQM, the time evolution operator depends on the state it acts on. As a result, Eqs. (\ref{eq:preSimple}) and (\ref{eq:postSimple})
become
\begin{equation}
p_{i\rightarrow f}^{{\rm NL}}=\left|\left\langle f|\mathcal{U}_{i}i\right\rangle \right|^{2},\qquad p_{i\leftarrow f}^{{\rm NL}}\propto\left|\left\langle i|\mathcal{U}_{f}^{\dagger}f\right\rangle \right|^{2},\label{eq:prepostSimpleNL}
\end{equation}
where, under some non-linear dynamics,  $\left|\mathcal{U}_{i}i\right\rangle $ is the time-evolved
$\left|i\right\rangle $ and $\left|\mathcal{U}_{f}^{\dagger}f\right\rangle $
is the backwards time-evolved $\left|f\right\rangle $. The superscript NL explicitly indicates that NLQM is being used. Moreover, the proportionality sign in $p_{i\leftarrow f}^{{\rm NL}}$
follows from  $\sum_{f}\left|\left\langle i|\mathcal{U}_{f}^{\dagger}f\right\rangle \right|^{2}$ being not, in general, normalized to unity. $p_{i\rightarrow f}^{{\rm NL}}$
and $p_{i\leftarrow f}^{{\rm NL}}$ are not necessarily equal, and so Born's rule cannot be uniquely extended to NLQM.

\subsection{Ambiguity in the boundary condition driving the non-linear time evolution}

By extending Eq. (\ref{eq:condState}) to NLQM, we can extend Born's rule, given by Eq. (\ref{eq:expressionProb}), to NLQM. However, because NLQM is non-linear, a time-evolution operator doesn't exist.  Nonetheless, inspired by the state-dependent Heisenberg picture introduced in \cite{helouSN}, we will show that we can define a \emph{boundary-dependent time-evolution operator}, and that the choice of a boundary condition is the essential degree of freedom for extending Born's rule to NLQM.

For some theories in NLQM\footnote{As mentioned in the introduction, we will not investigate all possible non-linear theories. In particular we only consider theories with the form of Eq. (\ref{eq:NLSE}), and whose solution can be uniquely specified with one Dirichlet boundary condition.}, running time-evolution requires solving the non-linear Schroedinger equation which contains a linear term, $\hat{H}_{L}$, and a nonlinear term $\hat{V}_{{\rm NL}}$:
\begin{equation}
i\hbar\partial_{t}\left|\psi\right\rangle =\left(\hat{H}_{L}+\hat{V}_{{\rm NL}}\left(\psi(x,t)\right)\right)\left|\psi\right\rangle.\label{eq:NLSE}
\end{equation}
$\hat{V}_{{\rm NL}}\left(\psi(x,t)\right)$ is a shorthand for a non-linear potential that  depends on the wavefunction $\ket{\psi\left(t\right)}$ expressed in some (possibly multi-dimensional) basis $x$. For instance, the Schroedinger-Newton equation for a single non-relativistic particle of mass $m$ interacting with its own gravitational field is given by
\begin{equation}
i\hbar\partial_{t}\left|\psi\right\rangle =\left(\hat{H}_{L}-m^2 G\int d^{3}\bm{x}\frac{\left|\psi\left(\bm{x},t\right)\right|^{2}}{\left|\hat{\bm{x}}-\bm{x}\right|}\right)\left|\psi\right\rangle,
\end{equation}
where $\psi\left(\bm{x},t\right)$ is the state $\ket{\psi}$ expressed in the position basis $\ket{\bm x}$. 

We chose to write the nonlinear Schroedinger equation in the form of Eq. (\ref{eq:NLSE}) to illustrate its similarity to the standard Schroedinger equation. Once we specify the boundary conditions, we can have Eq. (\ref{eq:NLSE}) be formally equivalent to a linear Schroedinger equation. We will assume that a single Dirichlet boundary condition is sufficient to solve Eq. (\ref{eq:NLSE}), and that its solution with the boundary condition $\ket{\psi(T)}=\ket{\phi}$ is $\ket{\varphi(t)}$. Consequently, the linear Schroedinger equation
\begin{equation}
i\hbar\partial_{t}\left|\psi\right\rangle =\left(\hat{H}_{L}+\hat{V}_{{\rm NL}}\left(\varphi(x,t)\right)\right)\left|\psi\right\rangle
\label{eq:equivalentlinearSE}
\end{equation}
is formally identical to Eq. (\ref{eq:NLSE}) with the boundary-condition $\ket{\psi(T)}=\ket{\phi}$. Heuristically, in the context of Eq. (\ref{eq:equivalentlinearSE}), we can interpret $\varphi(x,t)$ as a time-dependent classical drive. Eq. (\ref{eq:equivalentlinearSE}) has a time-evolution operator associated with it, which we denote by $\hat{U}_{\phi(T)}$. The subscript is to emphasize that the time-evolution operator is associated with the boundary condition $\ket{\psi(T)}=\ket{\phi}$. We can now write the solution to Eq. (\ref{eq:NLSE}) as
\begin{equation}
\left|\psi\left(t\right)\right\rangle =\hat{U}_{\phi\left(T\right)}\left(t,T\right)\left|\phi\right\rangle.
\end{equation}

%Heuristically,  $\hat{U}_{\phi(T)}$ can be understood as an all-powerful experimentalist's realization of  Eq. (\ref{eq:NLSE}) with sQM. The all-powerful experimentalist first solves Eq. (\ref{eq:NLSE}) with the appropriate boundary condition, and then, with their universal (linear) quantum simulation toolbox, constructs a Hamiltonian, with a time-dependent classical drive $\varphi(x,t)$, that is identical to the Hamiltonian in Eq. (\ref{eq:equivalentlinearSE}). Consequently, since Eq. (\ref{eq:equivalentlinearSE}) is equivalent to  Eq. (\ref{eq:NLSE}) with the boundary condition $\ket{\psi(T)}=\ket{\phi}$, they've realized Eq. (\ref{eq:NLSE}).
%Though, more practically, he doesn't need to do so, he can just simualte one setup at a time

We are ready to present the extension of Eqs. (\ref{eq:expressionProb}) and (\ref{eq:condState}) to NLQM:
\begin{equation}
p^{\rm NL}\left(\alpha,\beta\right|A_a, B_b)= \frac{1}{\mathcal{N}} \left\langle \fin{\alpha,\beta}|\fin{\alpha,\beta}\right\rangle ,\label{eq:expressionProbNL}
\end{equation}
where if Alice and Bob's measurement events are spacelike separated
\begin{equation}
\left|\Psi_{c|\alpha,\beta}\right\rangle =\hat{P}_{\beta}\hat{A}_{\phi^{A}_{2}\left(T^A_{2}\right)}\left(t_{2},t_{1}\right) \hat{B}_{\phi^{B}_{2}\left(T^B_{2}\right)}\left(t_{2},t_{1}\right) \hat{P}_{\alpha}\hat{U}_{\phi_{1}\left(T_{1}\right)}\left(t_{1},t_{0}\right)\left|\Psi_{{\rm ini}}\right\rangle ,
\label{eq:condStateNL}
\end{equation}
and $\mathcal{N}=\sum_{\alpha,\beta}\braket{\Psi_{c|\alpha,\beta}|\Psi_{c|\alpha,\beta}}$
ensures that $\sum_{\alpha,\beta}p^{{\rm NL}}\left(\alpha,\beta\right|A_a, B_b)$ is normalized to unity. Moreover, $\hat{U}_{\phi_{1}\left(T_{1}\right)}\left(t_{1},t_{0}\right)$ is the time-evolution operator from $t_0$ till $t_1$ and is associated with the boundary $\ket{\psi(T_1)}=\ket{\phi_1}$.  
$\hat{A}_{\phi^{A}_{2}\left(T_{2}\right)}\left(t_{2},t_{1}\right)$ ($\hat{B}_{\phi^{B}_{2}\left(T_{2}\right)}\left(t_{2},t_{1}\right)$) is the time-evolution operator associated with the boundary condition $\ket{\psi(T^A_2)}=\ket{\phi^A_{2}}$ ($\ket{\psi(T^B_2)}=\ket{\phi^B_{2}}$),  and acts on Alice's (Bob's) particle from $t_1$ till $t_2$. The time-evolution of Alice and Bob's joint system from $t_1$ till $t_2$ is separable because Alice's particle's future light-cone at $t_1$ does not overlap with Bob's particle's past light-cone at $t_2$. As a result, their total interaction  Hamiltonian, which includes contributions from the linear Hamiltonian $\hat{H}_L$ and from the non-linear potential $\hat{V}_{\rm NL}$, must be zero.
Note that, by construction, Eq. (\ref{eq:condStateNL}) recovers the predictions of sQM when the non-linearity $\hat{V}_{\rm{NL}}$ vanishes.

If the events A and B were time-like separated, then there are numerous schemes for extending the time evolution of Alice and Bob's particles to NLQM. 
Call the solutions  to Eq. (\ref{eq:NLSE}) with the boundary conditions  $\ket{\psi(T^A_2)}=\ket{\phi^A_{2}}$,  $\ket{\psi(T^B_2)}=\ket{\phi^B_{2}}$ and  $\ket{\psi(T^{AB}_2)}=\ket{\phi^{AB}_{2}}$ as $\ket{\varphi^{A}(t)}$, $\ket{\varphi^B(t)}$ and $\ket{\varphi^{AB}(t)}$ respectively.
We can then write the non-linear potential $\hat{V}_{\rm NL}$ in Eq. (\ref{eq:equivalentlinearSE}) in the following general way:
\begin{equation}
\hat{V}_{{\rm NL}}=\hat{V}_{A}\left(\varphi^{A}\left(x,t\right)\right)\otimes\hat{I}_B+\hat{I}_A\otimes\hat{V}_{B}\left(\varphi^{B}\left(x,t\right)\right)+\hat{V}_{{\rm int}}\left(\varphi^{{\rm AB}}\left(x,t\right)\right),
\end{equation}
where $\hat{V}_{A}$ ($\hat{V}_{B}$) is the free non-linear Hamiltonian acting on Alice's (Bob's) particle, and $\hat{V}_{\rm int}$ is the non-linear interaction potential.
However, we find it difficult to justify why each term in $\hat{V}_{\rm NL}$ would be generated by a different boundary condition when Alice and Bob's particles are allowed to directly communicate and interact. 
We will impose  $\phi_{2}^{A}=\phi_{2}^{B}=\phi_{2}^{AB}$ and $T_{2}^{A}=T_{2}^{B}=T_{2}^{AB}$ when the measurement events A and B are timelike separated.
We summarize our chosen form of $\left|\Psi_{c|\alpha,\beta}\right\rangle$ by
\begin{equation*}
\left|\Psi_{c|\alpha,\beta}\right\rangle =\begin{cases}
\hat{P}_{\beta}\hat{U}_{\phi_{2}\left(T_{2}\right)}\left(t_{2},t_{1}\right)\hat{P}_{\alpha}\hat{U}_{\phi_{1}\left(T_{1}\right)}\left(t_{1},t_{0}\right)\left|\Psi_{{\rm ini}}\right\rangle  & \mbox{A \& B timelike,}\\
\hat{P}_{\beta}\hat{A}_{\phi_{2}^{A}\left(T_{2}^{A}\right)}\left(t_{2},t_{1}\right)\hat{B}_{\phi_{2}^{B}\left(T_{2}^{B}\right)}\left(t_{2},t_{1}\right)\hat{P}_{\alpha}\hat{U}_{\phi_{1}\left(T_{1}\right)}\left(t_{1},t_{0}\right)\left|\Psi_{{\rm ini}}\right\rangle  & \mbox{A \& B spacelike.}
\end{cases}
\end{equation*}

The introduction of arbitrary boundary conditions $\ket{\phi_1}$ at $T_1$ and $\ket{\phi_2}$ at $T_2$ might appear artificial, but isn't. Each formulation of quantum mechanics predicts different boundary conditions after a measurement. For instance, in Eq. (\ref{eq:NLSE}),  an interpretation of quantum mechanics with wavefunction collapse  states that $\ket{\phi_1} = \ket{\alpha}$ and $T_1=t_1$, while the Everett interpretation states that $\ket{\phi_1}$ is the initial state of the universe and $T_1$ is when the universe began. Refer to section 4.1 for more details. In sQM, we do not have to worry if and how the wavefunction collapses because the time-evolution operator is well-defined independently of the wavefunction it acts on. However, in NLQM, each boundary condition generates a different time-evolution operator, and so how we formulate quantum mechanics matters in NLQM.

%$p\left(\alpha,\beta\right)$ can be extended to uncountably many ways in NLQM. For instance, we can exploit the fact that for any projection operator $\hat{P}^{\dagger}\hat{P}=\hat{P}$, and that $\hat{I}$ can be decomposed as the product of any unitary operator and its hermitian conjugate (e.g. $\hat{I}=\hat{U}_{1}^{\dagger}\hat{U}_{1}$). Each unitary operator, when extended to NLQM, can act on the state to its left or to its right. However, a lot of these prescriptions are meaningless because they aren't generated from $\Psi_c$ and so it is difficult for them to have an interpretation. Stated differently, it is like we generalize the bra $\Psi_c$ in one way and the ket $\Psi_c$ in a different way. Consequently, the expression for Born's rule won't be symmetric. This also doesn't make sense because the state-dependent Heisenberg picture is generated from a symmetric expectation value!

\subsection{Extending the formalism to relativistic quantum mechanics}

We can rigorously study superluminal communication only in quantum field theory, where the total Hamiltonian consists of free and interaction (between different fields)  energy densities
\begin{equation}
\hat{H}=\int d^{3}\bm{x}\hat{H}_{0}\left(\bm{x}\right)+\int d^{3}\bm{x}\hat{H}_{{\rm int}}\left(\bm{x}\right).
\end{equation}
Assigning spatial locations for quantum degrees of freedom is crucial for placing constraints on $\hat{H}$ to ensure that it is causal. Let $\hat{\mathcal H}_{\rm int}$ be the interaction energy density in an interaction picture with respect to $\int d^{3}\bm{x}\hat{H}_{0}\left(\bm{x}\right)$, then $\hat{\mathcal H}_{\rm int}$ commutes over spacelike distances \cite{duncan2012the}
\begin{equation}
\left[\hat{\mathcal H}_{{\rm int}}\left(t_x, \bm x \right),\,\hat{\mathcal H}_{{\rm int}}\left(t_y, \bm y\right)\right]=0,\qquad  c\left(t_{\bm{x}}-t_{\bm{y}}\right)^{2}-\left|\bm{x}-\bm{y}\right|^{2}<0.
\end{equation}
%The ordering of $t_1$ and $t_2$,  which depends on the reference frame that the experiment is viewed in, won't affect our results.

We generalize $\hat{H}$ to include a dependence on a wavefunction:
\begin{equation}
\hat{H}^{\rm NL}(t)=\int d^{3}\bm{x}\hat{H}_{0}\left(\bm{x},\left|\Psi_{\Phi\left(T\right)}\left(t\right)\right\rangle \right)+\int d^{3}\bm{x}\hat{H}_{{\rm int}}\left(\bm{x},\left|\Psi_{\Phi\left(T\right)}\left(t\right)\right\rangle \right),
\label{eq:nonlinearHQFT}
\end{equation}
where $\left|\Psi_{\Phi\left(T\right)}\left(t\right)\right\rangle$ is the solution to the non-linear Schroedinger equation 
\begin{equation}
i\hbar\partial_{t}\left|\Psi\left(t\right)\right\rangle =\left(\int d^{3}\bm{x}\hat{H}_{0}\left(\bm{x},\left|\Psi\left(t\right)\right\rangle \right)+\int d^{3}\bm{x}\hat{H}_{{\rm int}}\left(\bm{x},\left|\Psi\left(t\right)\right\rangle \right)\right)\left|\Psi\left(t\right)\right\rangle
\label{eq:QFTNLSE}
\end{equation}
with the boundary condition $\left|\Psi\left(t=T\right)\right\rangle =\left|\Phi\right\rangle$. We further generalize $\hat{H}^{{\rm NL}}$ by allowing for  different boundary conditions at each location
 \begin{equation}
\hat{H}^{{\rm NL}}(t)=\int d^{3}\bm{x}\hat{H}_{0}\left(\bm{x},\left|\Psi_{\Phi_{T}\left(\bm{x}\right)}(t)\right\rangle \right)+\int d^{3}\bm{x}\hat{H}_{{\rm int}}\left(\bm{x},\left|\Psi_{\Phi_{T}\left(\bm{x}\right)}\left(t\right)\right\rangle \right),
\label{eq;HNLQFT}
 \end{equation}
where  $\left|\Psi_{\Phi_{T}\left(\bm{x}\right)}\left(t\right)\right\rangle$ is the solution to Eq. (\ref{eq:QFTNLSE}) with boundary condition $\left|\Psi\left(t=T(\bm x)\right)\right\rangle =\left|\Phi(\bm x)\right\rangle$. 
 
The relativistic non-linear generalization of $\left|\Psi_{c|\alpha,\beta}\right\rangle$ in Eq. (\ref{eq:condStateComplicated}) is 
\begin{equation}
\left|\Psi_{c|\alpha,\beta}\right\rangle =\hat{P}_{\beta}\hat{U}_{\Phi^{(1)}_{T}\left(\bm{x}\right)}\left(t_{2},t_{1}\right)\hat{P}_{\alpha}\hat{U}_{\Phi^{(0)}_{T}\left(\bm{x}\right)}\left(t_{1},t_{0}\right)\left|\Psi_{{\rm ini}}\right\rangle,
\end{equation}
where $\bm{x} \in \mathbb{R}^{3}$, and $\hat{U}^{(0)}_{\Phi_{T}\left(t,\bm{x}\right)}$ ($\hat{U}^{(1)}_{\Phi_{T}\left(t,\bm{x}\right)}$) is the time-evolution operator associated with the boundary condition $\left|\Psi\left(t=T^{(0)}(\bm x)\right)\right\rangle =\left|\Phi^{(0)}(\bm x)\right\rangle$ ($\left|\Psi\left(t=T^{(1)}(\bm x)\right)\right\rangle =\left|\Phi^{(1)}(\bm x)\right\rangle$).

%To simplify the notation, we've identified the range of $t$ by the period of time that the time-evolution operator acts on. For instance, we associate  $\hat{U}_{\Phi_{T}\left(t,\bm{x}\right)}\left(t_{2},t_{1}\right)$ with time evolution between $t_1$ and $t_2$, and $\Phi_{T}\left(t,\bm{x}\right)$ specifies boundary conditions for times $t_1 \leq t \leq t_2$.
%we allow for changes in boundary states even between measurements, because the outcome of a measurement might not be known until after some delay time. It doesn't have to respond instantaneously to measurements. For that we will have an extra time dependence which is when the time evolution operator acts on.
 
\section{The no-signaling condition in NLQM}

As explained in section 2.1, Alice cannot communicate with Bob superluminally if $p^{\rm NL}(\beta|A_a, B_b)=\sum_{\alpha}p^{{\rm NL}}\left(\alpha,\beta\right|A_a, B_b)$ is independent of Alice's choice of measurement basis $A_a$. The normalization factor in $p^{{\rm NL}}\left(\alpha,\beta\right|A_a, B_b)$ (which we've shown explicitly in Eq. (\ref{eq:expressionProbNL})) won't affect our analysis and can be safely ignored for the remainder of this article\footnote{If the unnormalized $p^{\rm NL}(\beta|A_a, B_b)$ is independent of the basis $A_a$ for all $\beta$, then its normalization, $\sum_{\beta}p^{{\rm NL}}\left(\beta|A_a,B_b\right)$,  will also be independent of $A_a$. Moreover, it is obvious when the normalization could help: $p^{\rm NL}(\beta|A_a, B_b)$ is of the form $\left(\sum_{\alpha}f\left(\alpha\right)\right)g\left(\beta\right)$ where $f$ depends only on $\alpha$ and $g$ only on $\beta$. If such a scenario occurs, we will mention that the normalization eliminates the dependence of  $p^{\rm NL}(\beta|A_a, B_b)$ on $A_a$.}.
%We also offer a more general argument in the footnote\footnote{A non-trivial measurement apparatus has at least two outcomes, so both the numerator and denominator (which is the normalization) of $p^{{\rm NL}}\left(\beta\right|A_a, B_b)$ consist of a sum of at least two terms. The only way that $\mathcal{N}$ gets rid of the dependence of the numerator on the basis $A_a$ is if each term is of the form $A(\alpha) B\left(\beta\right)$, where $B\left(\beta\right)$ is a general function that depends only $\beta$, and $A(\alpha)$ a general function that only depends on $\alpha$. In such a case, we'd have 
%\begin{equation}
%p^{{\rm NL}}\left(\beta\right)=\frac{\left(\sum_{\alpha}A\left(\alpha\right)\right)B\left(\beta\right)}{\left(\sum_{\alpha}A\left(\alpha\right)\right)\left(\sum_{\beta}B\left(\beta\right)\right)}=\frac{B\left(\beta\right)}{\sum_{\beta}B\left(\beta\right)}.
%\end{equation}
%However, such a separable expression is not in general possible, because the initial state can be chosen to be anything, including entangled states, and the non-linear dynamics are arbitrary.
%}.

Similarly to how we derived Eq. (\ref{eq:pABLinSimple}), $p^{\rm NL}(\alpha,\beta|A_a, B_b)$ can be simplified to (ignoring the irrelevant normalization factor)
\begin{equation}
p^{{\rm NL}}\left(\alpha,\beta|A_a, B_b\right)=\braOket{\ini}{\hat{U}_{\phi_{1}\left(T_{1}\right)}^{\dagger}(t_1,t_0)\hat{B}_{\phi^B_{2}\left(T^B_{2}\right)}^{\dagger}\left(t_{2},t_{1}\right)\hat{P}_{\beta}\hat{B}_{\phi^B_{2}\left(T^B_{2}\right)}\left(t_{2},t_{1}\right)\hat{P}_{\alpha}\hat{U}_{\phi_{1}\left(T_{1}\right)}(t_1,t_0)}{\ini},
\label{eq:pabNLExpr}
\end{equation}
where we have used Eqs. (\ref{eq:PaPbcomm}), (\ref{eq:expressionProbNL}) and (\ref{eq:condStateNL}).
Before we perform a general analysis for arbitrary boundary states $\phi_1$, $\phi^A_2$ and $\phi^B_2$, we provide some examples.

\subsection{Some example formulations}

An interpretation that states that the wavefunction collapses after a measurement predicts
\begin{equation*}
p_{{\rm collapse}}^{{\rm NL}}\left(\alpha,\beta|A_a, B_b\right)=\braket{\hat{U}_{\ini\left(t_{0}\right)}^{\dagger}(t_{1},t_{0})\hat{B}_{\phi_{\alpha}\left(t_{2}\right)}^{\dagger}\left(t_{2},t_{1}\right)\hat{P}_{\beta}\hat{B}_{\phi_{\alpha}\left(t_{2}\right)}\left(t_{2},t_{1}\right)\hat{P}_{\alpha}\hat{U}_{\ini\left(t_{0}\right)}(t_{1},t_{0})},
\end{equation*}
where the expectation value is taken over $\ket{\ini}$ and
$\left|\phi_{\alpha}\right\rangle \equiv\hat{P}_{\alpha}\hat{U}_{\ini\left(t_{0}\right)}(t_{1},t_{0})\left|{\ini}\right\rangle$.
When we calculate $p^{{\rm NL}}_{\rm collapse}\left(\beta|A_a, B_b\right)$, we have to sum over $\alpha$ but since $\phi_\alpha$ depends on $\alpha$, the sum doesn't solely apply on $\hat{P}_\alpha$:
\begin{equation*}
p_{{\rm collapse}}^{{\rm NL}}\left(\beta|A_{a},B_{b}\right)=\braket{\hat{U}_{\ini\left(t_{0}\right)}^{\dagger}(t_{1},t_{0})\sum_{\alpha}\left(\hat{B}_{\phi_{\alpha}\left(t_{2}\right)}^{\dagger}\left(t_{2},t_{1}\right)\hat{P}_{\beta}\hat{B}_{\phi_{\alpha}\left(t_{2}\right)}\left(t_{2},t_{1}\right)\hat{P}_{\alpha}\right)\hat{U}_{\ini\left(t_{0}\right)}(t_{1},t_{0})}.
\end{equation*}
Consider another choice for Alice's measurement basis: $A_d$,  corresponding to an observable with eigenstates $\left|\delta\right\rangle$ and projection operators $\hat{D}_\delta$, then
\begin{equation*}
p_{{\rm collapse}}^{{\rm NL}}\left(\beta|A_{d},B_{b}\right)=\braket{\hat{U}_{\ini\left(t_{0}\right)}^{\dagger}(t_{1},t_{0})\sum_{\delta}\left(\hat{B}_{\varphi_{\delta}\left(t_{2}\right)}^{\dagger}\left(t_{2},t_{1}\right)\hat{P}_{\beta}\hat{B}_{\varphi_{\delta}\left(t_{2}\right)}\left(t_{2},t_{1}\right)\hat{D}_{\delta}\right)\hat{U}_{\ini\left(t_{0}\right)}(t_{1},t_{0})},
\end{equation*}
where $\left|\varphi_{\delta}\right\rangle \equiv\hat{D}_{\delta}\hat{U}_{\ini\left(t_{0}\right)}(t_{1},t_{0})\left|{\ini}\right\rangle$.
In general, $p_{{\rm collapse}}^{{\rm NL}}\left(\beta|A_{a},B_{b}\right)$ and  $p_{{\rm collapse}}^{{\rm NL}}\left(\beta|A_{d},B_{b}\right)$ aren't equal and so a formulation based on immediate wavefunction collapse violates the no-signaling condition. It also violates another tenet of special relativity: the statistics of measurement outcomes is not the same in all reference frames. Refer to the Appendix for more details.

On the other hand, a formulation of quantum mechanics in which collapse doesn't occur, such as the many-worlds interpretation, states 
\begin{equation}
p^{{\rm NL}}_{\rm M.W.}\left(\alpha,\beta|A_a, B_b\right)=\braket{\hat{U}_{\Phi_{\rm ini}\left(t_{\rm ini}\right)}^{\dagger}(t_1,t_0)\hat{B}_{\Phi_{\rm ini}\left(t_{\rm ini}\right)}^{\dagger}\left(t_{2},t_{1}\right)\hat{P}_{\beta}\hat{B}_{\Phi_{\rm ini}\left(t_{\rm ini}\right)}\left(t_{2},t_{1}\right)\hat{P}_{\alpha}\hat{U}_{\Phi_{\rm ini}\left(t_{\rm ini}\right)}(t_1,t_0)},
\label{eq:pABMW}
\end{equation}
where the expectation value is taken over $\ket{\ini}$, $t_{\rm ini}$ is when the universe began and $\ket{\Phi_{\rm ini}}$ is the initial state of the universe and so is independent of $\alpha$ and $\beta$. When calculating $p^{{\rm NL}}_{\rm M.W.}\left(\beta|A_a, B_b\right)$, the sum over $\alpha$  can be directly applied  on $\hat{P}_{\alpha}$ resulting in the identity operator, and so many-worlds does \emph{not} violate the no-signaling condition.
%However, the initial state of the universe is unknown and so Eq. (\ref{eq:pABMW}) is not a practical way to analyze an experiment. 
In the case of fundamental semi-classical gravity,  Eq. (\ref{eq:pABMW}) has already been ruled out \cite{pageArticle}.

In section 3.2, we discussed the prescriptions pre-selection and post-selection in the context of a single measurement. For the multiple measurements setup shown in Fig. \ref{fig:setup_measurements}, pre-selection takes $\phi_1$ and $\phi_2$ to be the initial state of the experiment $\left|\Psi_{{\rm ini}}\right\rangle$ and $T_2 = T_1 = t_0$. Post-selection takes $\phi_1$ and $\phi_2$ to be the final state of the experiment $\left|\alpha,\beta\right\rangle$ and $T_1 = T_2 = t_2$. 
Post-selection violates the no-signaling condition because both $\phi_1$ and $\phi_2$ depend on the measurement outcomes $\alpha$ and $\beta$. 
Pre-selection doesn't violate the no-signaling condition. However, although \cite{helouSN} treated it as a phenomenological prescription, it is equivalent to the Everett interpretation\footnote{
Choosing $\left|\phi_{1}\right\rangle$ and $\left|\phi_{2}\right\rangle$ to be the initial states of an experiment is not a well-defined procedure.
Consider again the setup shown in Fig. \ref{fig:setup_measurements}, where Charlie prepared $\left|\Psi_{{\rm ini}}\right\rangle$. He must have manipulated some state, which we call $\left|\Psi_{{\rm ini}}^{'}\right\rangle $, to prepare $\left|\Psi_{{\rm ini}}\right\rangle $. 
If we choose $\left|\Psi_{{\rm ini}}^{'}\right\rangle$ to be the initial state of the experiment, then pre-selection predicts that $\left|\phi_{1}\right\rangle =\left|\phi_{2}\right\rangle =\left|\Psi_{{\rm ini}}^{'}\right\rangle$.
This argument could be repeated back to the initial state of the universe. As a result, pre-selection seems to be equivalent to the Everett interpretation.}.

%Moreover, the particles evolve backwards-in-time (as the boundary conditions are placed at the end of the experiment rather than at the beginning), while in experiments time seems to flow forward in time.

\subsection{A general analysis}

From Eq. (\ref{eq:pabNLExpr}), we calculate $p^{\rm NL}(\beta|A_a, B_b)$ to be
\begin{equation*}
p^{{\rm NL}}\left(\beta|A_a, B_b\right)=\braket{\sum_{\alpha}\left(\hat{U}_{\phi_{1}\left(T_{1}\right)}^{\dagger}(t_1,t_0)\hat{B}_{\phi^B_{2}\left(T^B_{2}\right)}^{\dagger}(t_2,t_1)\hat{P}_{\beta}\hat{B}_{\phi^B_{2}\left(T^B_{2}\right)} (t_2,t_1)  \hat{P}_{\alpha}\hat{U}_{\phi_{1}\left(T_{1}\right)} (t_1,t_0) \right)},
\end{equation*}
and $p^{{\rm NL}}\left(\alpha|A_a, B_b\right)$ to be 
\begin{equation*}
p^{{\rm NL}}\left(\alpha|A_a, B_b\right)=\braket{\sum_{\beta}\left(\hat{U}_{\phi_{1}\left(T_{1}\right)}^{\dagger}(t_1,t_0)\hat{B}_{\phi^B_{2}\left(T^B_{2}\right)}^{\dagger}(t_2,t_1)\hat{P}_{\beta}\hat{B}_{\phi^B_{2}\left(T^B_{2}\right)} (t_2,t_1)  \hat{P}_{\alpha}\hat{U}_{\phi_{1}\left(T_{1}\right)} (t_1,t_0) \right)},
\end{equation*}
where for both probabilities, the expectation value is taken over $\ket{\ini}$.
The no-signaling condition is violated if $p^{{\rm NL}}\left(\beta|A_a, B_b\right)$ depends on $A_a$ or if $p^{{\rm NL}}\left(\alpha|A_a, B_b\right)$ depends on $B_b$. 

Notice that if $\phi_1$ depends on $\alpha$ then $p^{{\rm NL}}\left(\beta|A_a, B_b\right)$ depends on $A_a$ and so $p^{{\rm NL}}\left(\beta|A_a, B_b\right) \neq p^{{\rm NL}}\left(\beta|B_b\right)$. Similarly, if $\phi_1$ depends on $\beta$ then $p^{{\rm NL}}\left(\alpha|A_a, B_b\right)$ depends on $B_b$. Consequently, $\phi_1$ must be independent of $\alpha$ and $\beta$. Similarly, $\phi^B_2$ must also be independent of $\alpha$ and $\beta$. On the other hand, $\phi^A_2$ is unconstrained, and so our analysis doesn't result in a unique prescription. 
Nonetheless, we find it difficult to justify why  $\phi_1$ and $\phi^B_2$ would be anything other than the initial state of the experiment or of the universe. If we choose all boundary states to be the initial state of the universe, then we recover the Everett interpretation. In the next section, we discuss another reasonable prescription for assigning boundary states.

\section{Causal-conditional: A sensible prescription that doesn't violate the no-signaling condition}

In this section, we propose and discuss a prescription, which we name \emph{causal-conditional}, for assigning boundary states to time-evolution operators in a way that doesn't violate the no-signaling condition.
The causal-conditional prescription updates the boundary states of degrees of freedom lying in the future light cone of a particular measurement.  We will be conservative and not explicitly assign a mechanism for this process, be it objective collapse or emergent behavior after the wavefunction branches. 
We only specify that the predictions of causal-conditional are mathematically equivalent to sQM with causal feedback following each measurement event.

To precisely explain the causal-conditional prescription, we will present, using the language of quantum field theory introduced in section 3.3, the quantum state of a general collection of degrees of freedom at an arbitrary time $t_{f}$. We'll assume that their initial state at time $t_{0}$ is $\left|\Psi_{{\rm ini}}\right\rangle$ and that $N$ measurements have occurred up to the final time $t_{f}$.
The (unnormalized) conditional state at $t_{f}$ is 
\begin{equation}
\left|\Psi_{c}\right\rangle =\hat{U}_{N}\hat{P}\left(t_{N},\bm{x}_{N}\right)\hat{U}_{N-1}...\hat{P}\left(t_{1},\bm{x}_{1}\right)\hat{U}_{0}\left|\Psi_{{\rm ini}}\right\rangle ,
\end{equation}
where $\hat{P}\left(t',\bm{y}\right)$ is a projection operator associated with a measurement at the spacetime location $\left(t',\bm{y}\right)$ and $\hat{U}_{i}$, for
$0\leq i\leq N$, is the time-evolution operator from $t_{i-1}$ till
$t_{i}$. After some explanation, we provide $\hat{U}_{i}$'s exact
expression in Eq. (\ref{eq:UiDefEq}). 

According to the causal-conditional prescription, a degree of freedom modifies the boundary condition that the non-linearity at its spacetime location depends on (as in Eq. (\ref{eq;HNLQFT})) when it receives information about a measurement outcome.
This information propagates along the future light cone of a measurement event.
Assume that for some $1\leq i\leq N$, a degree of freedom at $\bm{x}$ receives information, at times $s_{1}^{\left(i\right)},...,s_{m_{i}}^{\left(i\right)}$ between $t_{i-1}$ and $t_{i}$,
about $m_{i}$ measurement outcomes, then 
\begin{equation}
\hat{U}_{i}=\hat{U}_{\Phi_{T}^{\left(i\right)}\left(s_{m_{i}}^{\left(i\right)},\bm{x}\right)}\left(t_{i},s_{m_{i}}^{\left(i\right)}\right)...\hat{U}_{\Phi_{T}^{\left(i\right)}\left(s_{1}^{\left(i\right)},\bm{x}\right)}\left(s_{2}^{\left(i\right)},s_{1}^{\left(i\right)}\right)\hat{U}_{\Phi_{T}^{\left(i\right)}\left(t_{i-1},\bm{x}\right)}\left(s_{1}^{\left(i\right)},t_{i-1}\right).\label{eq:UiDefEq}
\end{equation}
Note that we have extended the definition of the boundary state $\left|\Phi\left(\bm{x}\right)\right\rangle $
to include a dependence on time: $\left|\Phi\left(t,\bm{x}\right)\right\rangle $.
Moreover, no measurements occurred before $t_{1}$ so $\hat{U}_{0}=\hat{U}_{\Phi_{T}^{\left(0\right)}\left(t_{0},\bm{x}\right)}\left(t_{1},t_{0}\right)$.

\begin{figure}
\centering\includegraphics[scale=0.4]{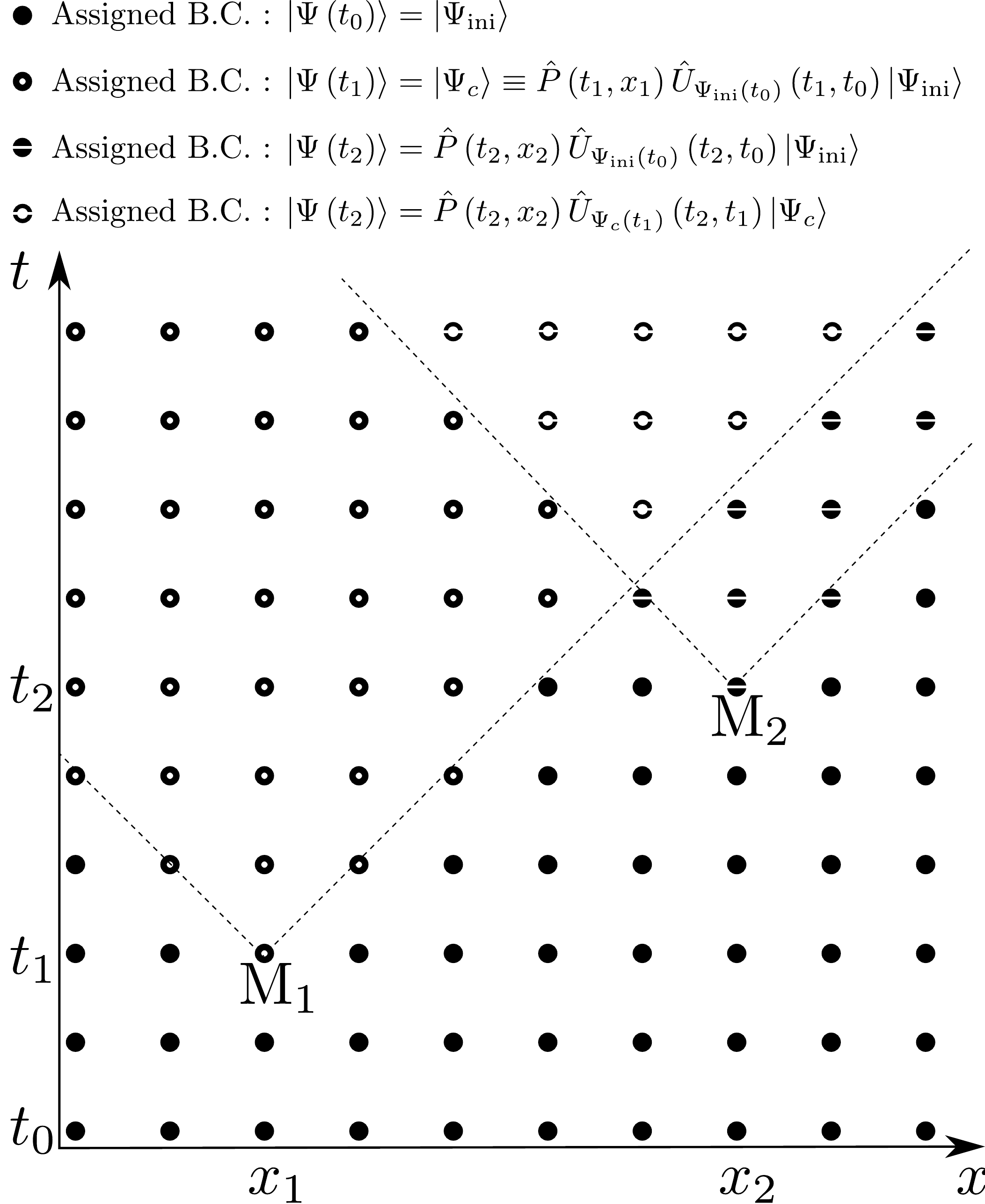}

\caption{\label{fig:causal_conditional_QFT_explanation}
Assignment of boundary conditions after two measurements according to the causal-conditional prescription.
$\mbox{M}_1$ and $\mbox{M}_2$ are two measurement events at spacetime locations $(t_1, x_1)$ and $(t_2, x_2)$, respectively, and the dashed lines show the light cones centered around each of them.
To keep the figure uncluttered, we work with a one-dimensional quantum field, and we have discretized space and time into 10 points each.
Each degree of freedom of the field is represented by a dot on the figure. 
How we fill the dot depends on what boundary condition (B.C.), which is indicated on the legend at the top of the figure, is assigned to the time-evolution of the wavefunction that the non-linear Hamiltonian at the spatial location of the dot depends on (see section 3.3 for more details).
Note that the initial state of the field is $\left|\Psi_{{\rm ini}}\right\rangle$.
}

\end{figure}

The causal-conditional prescription chooses the boundary states as
follows. For $t_{0}\le t<t_{1}$, no measurements have occurred, so $\left|\Phi^{\left(0\right)}\left(t,\bm{x}\right)\right\rangle =\left|\Psi_{{\rm ini}}\right\rangle $
for all $t$ and the boundary time is $T^{\left(0\right)}\left(\bm{x}\right)=t_{0}$.
For $1\leq i\leq N$ , $T^{\left(i\right)}\left(\bm{x}\right)=t_{i}$.
The $\left|\Phi^{\left(i\right)}\left(t,\bm{x}\right)\right\rangle$ are defined sequentially from $i=1$ till $i=N$:
\begin{eqnarray}
\left|\Phi^{\left(1\right)}\left(t,\bm{x}\right)\right\rangle  & \equiv & \hat{\mathcal{P}}_{\left(t,\bm{x}\right)}\left(t_{1},\bm{x}_{1}\right)\hat{U}_{\Phi_{T}^{\left(0\right)}\left(t,\bm{x}\right)}\left(t_{1},t_{0}\right)\left|\Psi_{{\rm ini}}\right\rangle ,\\
\left|\Phi^{\left(2\right)}\left(t,\bm{x}\right)\right\rangle  & \equiv & \mathcal{\hat{P}}_{\left(t,\bm{x}\right)}\left(t_{2},\bm{x}_{2}\right)\hat{U}_{\Phi_{T}^{\left(1\right)}\left(t,\bm{x}\right)}\left(t_{2},t_{1}\right)\left|\Phi^{\left(1\right)}\left(t,\bm{x}\right)\right\rangle ,\\ \nonumber
\vdots & \vdots & \vdots  \\ 
\left|\Phi^{\left(N\right)}\left(t,\bm{x}\right)\right\rangle  & \equiv & \mathcal{\hat{P}}_{\left(t,\bm{x}\right)}\left(t_{N},\bm{x}_{N}\right)\hat{U}_{\Phi_{T}^{\left(N-1\right)}\left(t,\bm{x}\right)}\left(t_{N},t_{N-1}\right)\left|\Phi^{\left(N-1\right)}\left(t,\bm{x}\right)\right\rangle ,
\end{eqnarray}
for all $t\geq t_{0}$ and where $\hat{\mathcal{P}}_{(t,\bm{x})}\left(t',\bm{y}\right)$
is $\hat{P}\left(t',\bm{y}\right)$ if $(t,\bm{x})$ lies in $(t',\bm{y})$
's future light cone, and the identity operator otherwise: 
\begin{equation}
\hat{\mathcal{P}}_{(t,\bm{x})}\left(t',\bm{y}\right)\equiv\hat{I}+\left(\hat{P}\left(t',\bm{y}\right)-\hat{I}\right)\theta\left(\Delta s^{2}\left\{ \left(t,\bm{x}\right),\left(t',\bm{y}\right)\right\} \right).
\end{equation}
$\theta(t)$ is the Heaviside function and $\Delta s^{2}\left\{ \left(t,\bm{x}\right),\left(t',\bm{y}\right)\right\} $
is the spacetime distance between events $\left(t,\bm{x}\right)$
and $(t',\bm{y})$. We illustrate the assignment of boundary states
after the first two measurements in Fig. \ref{fig:causal_conditional_QFT_explanation}. 

Finally, note that our scheme is similar to Adrian Kent's proposal in \cite{KentIdea}. He argued that if the non-linear time evolution depends only on local states, which are obtained by conditioning only on measurements in the past light cone of a degree of freedom, then superluminal communication is not possible. 

\subsection{An example}
\begin{figure}[h]
\begin{minipage}{18pc}
\includegraphics[width=18pc]{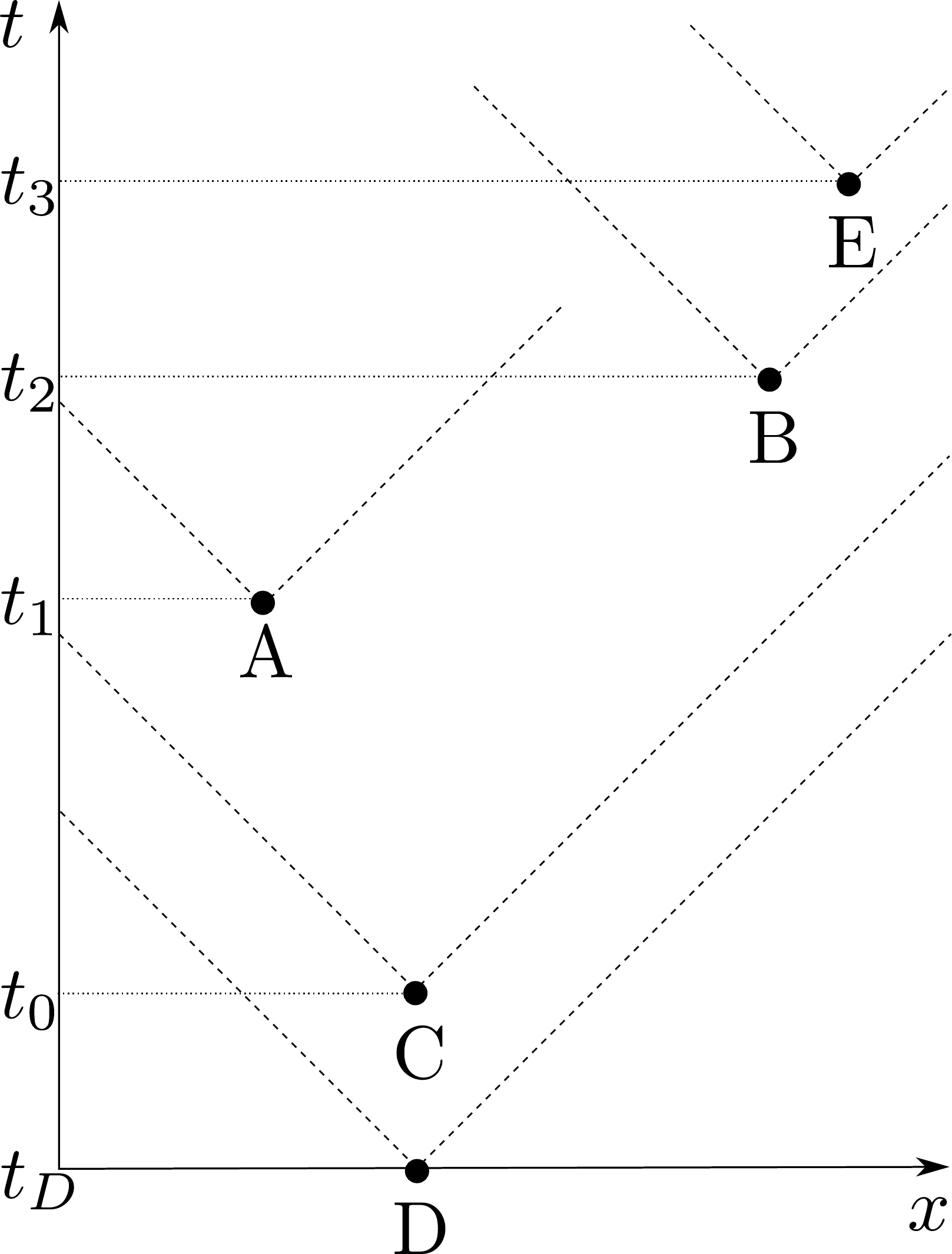}
\caption{\label{fig:Setup-For-condPre}A setup similar to that described by Fig. \ref{fig:setup_measurements}, but more elaborate. Event D is Dylan preparing the state $\left|\Psi_{{\rm ini}}^{'}\right\rangle $, Event $C$ is Charlie measuring the eigenstate $\left|\Psi_{{\rm ini}}\right\rangle $. Event A (B) describes Alice (Bob) measuring her (his) particles. Bob then sends his particle to be measured by Eve at event E. The dashed lines show the light cone centered around each event. }
\end{minipage}\hspace{2pc}%
\begin{minipage}{18pc}
\includegraphics[width=18pc]{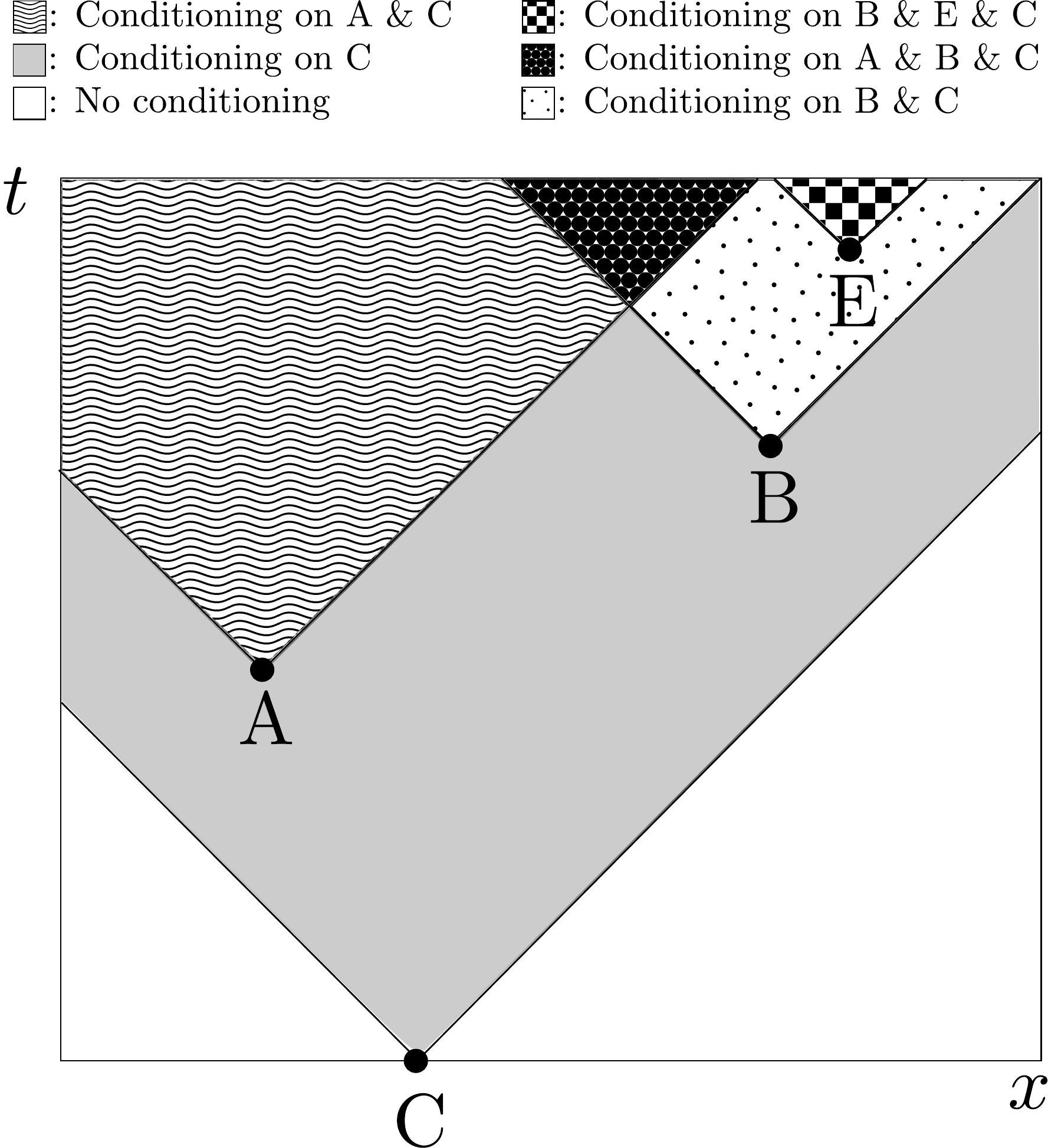}
\caption{\label{fig:stateDifferentRegions}Partioning of spacetime into different regions  according to which boundary state is associated with time evolution.  There are 4 measurement events: C, A, B and E, that we've arranged identically as in Fig. \ref{fig:Setup-For-condPre}. We didn't include Event D to limit clutter. The 4 events result in 6 regions. The boundary state associated with the non-linear time-evolution operator of each region is the time-evolved initial state of the experiment conditioned on measurement events presented in the legend at the top of the figure. }
\end{minipage} 
\end{figure}

Consider the setup shown in Fig. \ref{fig:Setup-For-condPre}, which is a more elaborate version of Fig. \ref{fig:setup_measurements}. 
The thought experiment now includes two additional parties: 
Dylan who prepares an ensemble of two particles in the state $\left|\Psi_{{\rm ini}}^{'}\right\rangle$ and then sends them to Charlie, and Eve who performs a measurement outside the future light cone of Alice on Bob's particle at time $t_3$.
We've added Dylan to demonstrate that we don't need to know $\left|\Psi_{{\rm ini}}^{'}\right\rangle$ to predict the distribution of outcomes for measurements lying in the future light cone of Dylan's measurement. We've added Eve to show that even for a complicated configuration of measurement events, our prescription does not violate the no-signaling condition.

We begin our analysis with the predictions of sQM for the final unnormalized state
of the experiment conditioned on Charlie, Alice, Bob and Eve measuring
$\gamma$, $\alpha$, $\beta$ and $\epsilon$ with corresponding measurement
eigenstates $\left|\Psi_{{\rm ini}}\right\rangle $, $\left|\alpha\right\rangle $,
$\left|\beta\right\rangle $ and $\left|\epsilon\right\rangle $, and corresponding measurement bases $C_c$, $A_a$, $B_b$ and $E_e$,
respectively:
\begin{eqnarray}
\left|\psi_{c}\right\rangle & = &  \hat{P}_{\epsilon}\hat{U}(t_3,t_2)\hat{P}_{\beta}\hat{U}(t_2,t_1)\hat{P}_{\alpha}\hat{U}(t_1,t_0)\hat{P}_{\gamma}\hat{U}(t_0,t_D)\left|\Psi_{{\rm ini}}^{'}\right\rangle.
\end{eqnarray}
The projection operators are
\begin{equation}
\hat{P}_{\gamma}=\left|\Psi_{{\rm ini}}\right\rangle \left\langle \Psi_{{\rm ini}}\right|,\quad\hat{P}_{\alpha}=\left|\alpha\right\rangle \left\langle \alpha\right|,\quad\hat{P}_{\beta}=\left|\beta\right\rangle \left\langle \beta\right|,\quad\hat{P}_{\epsilon}=\left|\epsilon\right\rangle \left\langle \epsilon\right|.
\end{equation}
The structure of $\left|\psi_{c}\right\rangle $ can be simplified by noticing that Alice's measurement's future light cone doesn't overlap with Bob and Eve's measurement events' past light cone. We obtain
\begin{eqnarray}
\left|\psi_{c}\right\rangle =\hat{P}_{\epsilon}\hat{A}\left(t_{3},t_{2}\right)\hat{E}\hat{P}_{\beta}\hat{A}\left(t_{2},t_{1}\right)\hat{B}\hat{P}_{\alpha}\hat{U}\hat{P}_{\gamma}\hat{V}\left|\Psi_{{\rm ini}}^{'}\right\rangle ,
\end{eqnarray}
where to keep the notation concise, we have made the following definitions
\begin{equation}
\hat{V}\equiv\hat{U}\left(t_{0},t_{D}\right);\qquad\hat{U}\equiv\hat{U}\left(t_{1},t_{0}\right),
\end{equation}
and $\hat{B}$ and $\hat{E}$ are the time-evolution operators for Bob and Eve's measured degrees of freedom from $t_1$ till $t_2$ and from $t_2$ till $t_3$, respectively.

According to the causal-conditional prescription, $\left|\psi_{c}\right\rangle $ extends to NLQM in the following way:
\begin{eqnarray}
\left|\psi_{c}\right\rangle  & = & \hat{P}_{\epsilon}\hat{A}_{\alpha\left(t_{1}\right)}\left(t_{3},t_{2}\right)\hat{E}_{\phi_{3}\left(t_{2}\right)}\hat{P}_{\beta}\hat{A}_{\alpha\left(t_{1}\right)}\left(t_{2},t_{1}\right)\hat{B}_{\Psi_{{\rm ini}}\left(t_{0}\right)}\hat{P}_{\alpha}\hat{U}_{\Psi_{{\rm ini}}\left(t_{0}\right)}\hat{P}_{\gamma}\hat{V}_{\Psi_{{\rm ini}}^{'}\left(t_{D}\right)}\left|\Psi_{{\rm ini}}^{'}\right\rangle \\
 & \propto & \hat{A}_{\alpha\left(t_{1}\right)}\left(t_{3},t_{2}\right) \hat{A}_{\alpha\left(t_{1}\right)}\left(t_{2},t_{1}\right) \hat{P}_{\epsilon}\hat{E}_{\phi_{3}\left(t_{2}\right)}\hat{P}_{\beta}\hat{B}_{\Psi_{{\rm ini}}\left(t_{0}\right)}\hat{P}_{\alpha}\hat{U}_{\Psi_{{\rm ini}}\left(t_{0}\right)}\left|\Psi_{{\rm ini}}\right\rangle,
 \label{eq:PsicNewPres}
\end{eqnarray}
where
\begin{equation}
\left|\phi_{3}\right\rangle =\hat{P}_{\beta}\hat{A}_{\Psi_{{\rm ini}}\left(t_{0}\right)}\left(t_{2},t_{1}\right)\hat{B}_{\Psi_{{\rm ini}}\left(t_{0}\right)}\hat{U}_{\Psi_{{\rm ini}}\left(t_{0}\right)}\left|\Psi_{{\rm ini}}\right\rangle.
\end{equation}
We have also used that Alice's particle doesn't interact with the second particle after $t_1$ and so $\hat{A}$ commutes with $\hat{B}$, $\hat{E}$, $\hat{P}_\beta$ and $\hat{P}_\epsilon$.
Bob's past light cone does not include Alice's measurement event, but includes Charlie's, so $\left|\Psi_{{\rm ini}}\right\rangle$ is the boundary state associated with Bob's particle's time-evolution operator $\hat{B}$.
Moreover, Eve's past light cone includes Bob's measurement event, so the conditional state $\ket{\phi_3}$ is the boundary state associated with $\hat{E}$.
Notice that $\hat{A}$ in $\ket{\phi_3}$ is associated with the boundary state $\Psi_{\rm ini}$. Refer to Fig. \ref{fig:stateDifferentRegions} for more details.

Eq. (\ref{eq:PsicNewPres}) doesn't violate the no-signaling condition and contains genuine non-linear time evolution, such as $\hat{U}_{\Psi_{{\rm ini}}\left(t_{0}\right)}\left|\Psi_{{\rm ini}}\right\rangle$. 
Moreover, notice that measurements within the past light cone of Charlie, like that of Dylan's, do not affect our analysis. 
Indeed, preparation events are always in the past light cone of the final measurements of an experiment because the measured particles' speed is upper bounded by the speed of light.
Consequently, experimentalists do not need to know about measurements occurring outside their experimental setup to calculate the predictions of the causal-conditional prescription.
%avoids the issue that pre-selection and the Everett interpretation suffer from, of trying to figure what states outside that of the experiment are.

We show that our proposed prescription does not violate the no-signaling
condition by looking at the marginal probabilities, $p\left(\alpha|C, \mathcal{B}\right)$,
$p\left(\beta|C, \mathcal{B}\right)$ and $p\left(\epsilon|C, \mathcal{B}\right)$,
conditioned on Charlie measuring $\left|\Psi_{{\rm ini}}\right\rangle $ and on the measurement bases $\mathcal{B}\equiv\left\{ C_{c},A_{a},B_{b},E_{e}\right\}$.
The probability of obtaining the measurement results $\alpha,\beta,\epsilon$,
and that Charlie measures $\left|\Psi_{{\rm ini}}\right\rangle $
is given by the norm of $\left|\Psi_{c}\right\rangle $:
\begin{eqnarray*}
 p^{{\rm NL}}\left(\alpha,\beta,\epsilon,C| \mathcal{B}\right) & = & \left|\braOket{\Psi_{{\rm ini}}}{\hat{V}_{\Psi_{{\rm ini}}^{'}\left(t_{D}\right)}}{\Psi_{{\rm ini}}^{'}}\right|^{2}  \times\\
 &  & \braOket{\Psi_{{\rm ini}}}{\hat{U}_{\Psi_{{\rm ini}}\left(t_{0}\right)}^{\dagger}\hat{B}_{\Psi_{{\rm ini}}\left(t_{0}\right)}^{\dagger}\hat{P}_{\beta}\hat{E}_{\phi_{3}\left(t_{2}\right)}^{\dagger}\hat{P}_{\epsilon}\hat{E}_{\phi_{3}\left(t_{2}\right)}\hat{P}_{\beta}\hat{B}_{\Psi_{{\rm ini}}\left(t_{0}\right)}\hat{P}_{\alpha}\hat{U}_{\Psi_{{\rm ini}}\left(t_{0}\right)}}{\Psi_{{\rm ini}}},
\end{eqnarray*}
%Note that this is already normalized to unity
%We have also simplified $p^{{\rm NL}}\left(\alpha,\beta,\epsilon, C \right)$ by noting that $\hat{P}_{\alpha}$, which corresponds to an event space-like separated from Bob and Eve's measurements, commutes with $\hat{U}_{2},$ $\hat{U}_{3}$, $\hat{P}_{\beta}$ and $\hat{P}_{\epsilon}$. 
We are interested in $p^{{\rm NL}}\left(\alpha,\beta,\epsilon|C, \mathcal{B}\right)$,
so we have to divide by 
\begin{equation}
p^{{\rm NL}}\left(C|\mathcal{B}\right)=\sum_{\alpha,\beta,\epsilon}p^{{\rm NL}}\left(\alpha,\beta,\epsilon,C|\mathcal{B}\right)=\left|\braOket{\Psi_{{\rm ini}}}{\hat{V}_{\Psi_{{\rm ini}}^{'}\left(t_{D}\right)}}{\Psi_{{\rm ini}}^{'}}\right|^{2}.
\end{equation}
We obtain
\begin{equation}
 p^{{\rm NL}}\left(\alpha,\beta,\epsilon|C, \mathcal{B}\right)  = 
 \braOket{\Psi_{{\rm ini}}}{\hat{U}_{\Psi_{{\rm ini}}\left(t_{0}\right)}^{\dagger}\hat{B}_{\Psi_{{\rm ini}}\left(t_{0}\right)}^{\dagger}\hat{P}_{\beta}\hat{E}_{\phi_{3}\left(t_{2}\right)}^{\dagger}\hat{P}_{\epsilon}\hat{E}_{\phi_{3}\left(t_{2}\right)}\hat{P}_{\beta}\hat{B}_{\Psi_{{\rm ini}}\left(t_{0}\right)}\hat{P}_{\alpha}\hat{U}_{\Psi_{{\rm ini}}\left(t_{0}\right)}}{\Psi_{{\rm ini}}}.
\end{equation}

Can Alice send signals to Bob or Eve, or vice versa? We first calculate $p^{{\rm NL}}\left(\beta|C, \mathcal{B}\right)$: 
\begin{eqnarray}
p^{{\rm NL}}\left(\beta|C, \mathcal{B}\right) & = & \braOket{\Psi_{{\rm ini}}}{\hat{U}_{\Psi_{{\rm ini}}\left(t_{0}\right)}^{\dagger}\hat{B}_{\Psi_{{\rm ini}}\left(t_{0}\right)}^{\dagger}\hat{P}_{\beta}\hat{B}_{\Psi_{{\rm ini}}\left(t_{0}\right)}\hat{U}_{\Psi_{{\rm ini}}\left(t_{0}\right)}}{\Psi_{{\rm ini}}},
\end{eqnarray}
which doesn't depend on $A_a$. Next, we calculate Eve's distribution of measurement results:
\begin{equation}
p^{{\rm NL}}\left(\epsilon|C, \mathcal{B}\right) = \braOket{\Psi_{{\rm ini}}}{\hat{U}_{\Psi_{{\rm ini}}\left(t_{0}\right)}^{\dagger}\hat{B}_{\Psi_{{\rm ini}}\left(t_{0}\right)}^{\dagger}\sum_{\beta}\left(\hat{P}_{\beta}\hat{E}_{\phi_{3}\left(t_{2}\right)}^{\dagger}\hat{P}_{\epsilon}\hat{E}_{\phi_{3}\left(t_{2}\right)}\hat{P}_{\beta}\right)\hat{B}_{\Psi_{{\rm ini}}\left(t_{0}\right)}\hat{U}_{\Psi_{{\rm ini}}\left(t_{0}\right)}}{\Psi_{{\rm ini}}},
\end{equation}
so Alice cannot send superluminal signals to Eve. Bob and Eve's measurement events are time-like separated so it is acceptable that they can communicate amongst each other. Finally, Bob and Eve cannot communicate to Alice superluminally because
\begin{eqnarray}
p^{{\rm NL}}\left(\alpha|C, \mathcal{B}\right) & = & \braOket{\Psi_{{\rm ini}}}{\hat{U}_{\Psi_{{\rm ini}}\left(t_{0}\right)}^{\dagger}\hat{P}_{\alpha}\hat{U}_{\Psi_{{\rm ini}}\left(t_{0}\right)}}{\Psi_{{\rm ini}}}
\end{eqnarray}
doesn't depend on $B_b$ and $E_e$.

\subsection{Proof that the causal-conditional prescription  doesn't violate the no-signaling condition}
We prove that the prescription discussed in this section does not violate the no-signaling condition. We first present a heuristic argument. 
The causal-conditional prescription is mathematically equivalent to linear quantum mechanics with causal feedback, and so doesn't violate the no-signaling condition. In particular, whenever a measurement occurs, the wavefunction that the non-linear potential depends on isn't modified instantaneously. Instead, a measurement transmits its outcome along its future light cone. Degrees of freedom that receive this information update their boundary state accordingly.

We now present a rigorous argument. Consider a general measurement configuration as viewed in some reference frame. The unnormalized conditional state after the final measurement is 
\begin{equation}
\left|\psi_{c}\right\rangle =\hat{U}_{1}\hat{P}_{1}\left(\alpha_{1}\right)...\hat{U}_{f}\hat{P}_{f}\left(\alpha_{f}\right)\left|{\rm ini}\right\rangle ,
\end{equation}
where $\ket{\rm ini}$ is the initial state of all degrees of freedom before the first measurement, $\hat{P}_i(\alpha_i)$ is the projection operator  (with outcome $\alpha_i$) associated with the $i$th measurement, and the $\hat{U}_1$, $\hat{U}_2$, ..., $\hat{U}_f$ are boundary-dependent time-evolution operators. 

Assume that Bob performs, at time $t_B$, one of these measurement. 
We will show that Bob's probability of measuring a particular outcome $\beta$, 
\begin{equation}
p\left(\beta|\Omega\right)=\sum_{\alpha_{1},...,\alpha_{f}}{\vphantom{\sum}}'  \braOket{{\rm ini}}{\hat{U}_{1}^{\dagger}\hat{P}_{1}\left(\alpha_{1}\right)...\hat{P}_{f-1}\left(\alpha_{f-1}\right)\hat{U}_{f}^{\dagger}\hat{P}_{f}\left(\alpha_{f}\right)\hat{U}_{f}\hat{P}_{f-1}\left(\alpha_{f-1}\right)...\hat{P}_{1}\left(\alpha_{1}\right)\hat{U}_{1}}{{\rm ini}},
\end{equation}
where $\Omega$ is the set of all chosen measurement bases, is independent of measurements after $t_B$, and outside Bob's measurement's past light cone. Note that the sum is over all measurement outcomes except Bob's.
All measurements occurring after $t_B$ do not matter because we can directly sum over them. 
Let's first sum over $\alpha_f$. We obtain
\begin{equation}
p\left(\beta|\Omega\right)=\sum_{\alpha_{1},...,\alpha_{f-1}}{\vphantom{\sum}}' 
\braOket{{\rm ini}}{\hat{U}_{1}^{\dagger}\hat{P}_{1}\left(\alpha_{1}\right)...\hat{U}_{f-1}^{\dagger}\hat{P}_{f-1}\left(\alpha_{f-1}\right)\hat{U}_{f-1}...\hat{P}_{1}\left(\alpha_{1}\right)\hat{U}_{1}}{{\rm ini}}
\end{equation}
because the final measurement does not lie in the past light cone of any other measurement, and so no time-evolution operator would depend on $\alpha_f$.
We can repeat this procedure for all other measurements events after $t_B$.

\begin{figure}
\centering\includegraphics[scale=0.4]{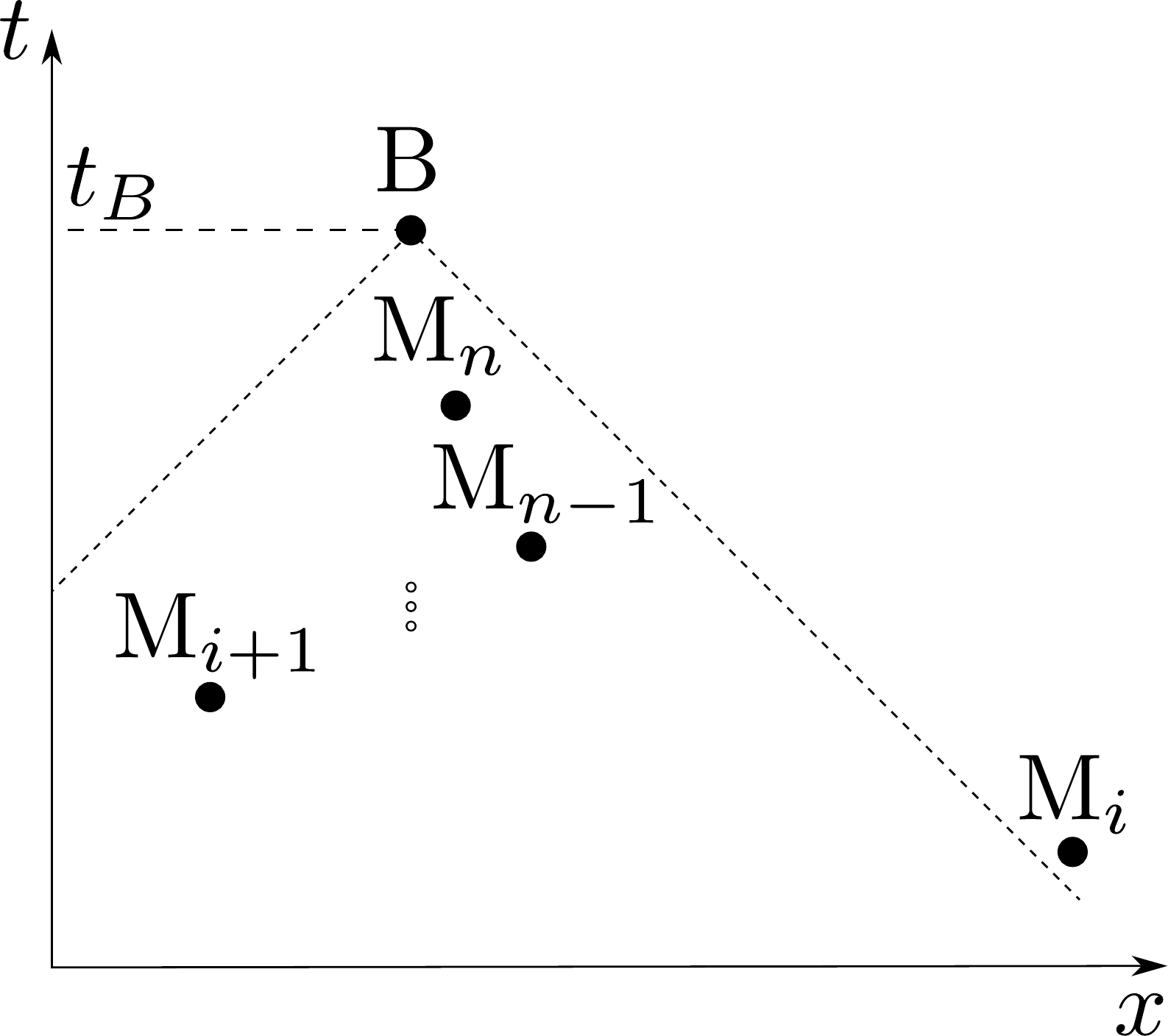}

\caption{\label{fig:general_argument}
A general configuration of measurements, labeled by $M_j$ where $ 1 \leq j \leq n$, occurring before an event B, which describes Bob performing a measurement.
The dashed lines shows the past light cone of event B. }

\end{figure}

For this part of the proof, we label the projection operator corresponding to Bob's measurement by $\hat{P}_\beta$ and  assume that $n$ measurements precede Bob's. 
Let $\tilde{\Omega}\subset\Omega$ be the set of measurements bases chosen by Bob and all experimentalists performing measurements before Bob. 
After summing over all the outcomes of all measurements performed after Bob's, we then obtain that
\begin{equation}
p\left(\beta|\tilde{\Omega}\right)=\sum_{\alpha_{1},...,\alpha_{n}}\braOket{{\rm ini}}{\hat{U}_{1}^{\dagger}\hat{P}_{1}\left(\alpha_{1}\right)...\hat{P}_{n}\left(\alpha_{n}\right)\hat{U}_{n+1}^{\dagger}\hat{P}_{\beta}\hat{U}_{n+1}\hat{P}_{n}\left(\alpha_{n}\right)...\hat{P}_{1}\left(\alpha_{1}\right)\hat{U}_{1}}{{\rm ini}}.
\label{eq:generalPB}
\end{equation}
Consider the measurement occurring closest to $t_B$, and that is outside Bob's measurement's past light cone, as shown in Fig. \ref{fig:general_argument}. Assume it corresponds to the $i$th measurement event, and so according to the causal-conditional prescription, the time-evolution operators $\hat{U}_{i+1}$, ..., $\hat{U}_{n+1}$ contain boundary terms dependent on  $\alpha_i$. Let's explicitly separate each of them into two components: $\hat{U}_j \equiv \hat{V}_j \hat{W}_j$ for $i+1 \leq j \leq n+1$ and where $\hat{V}_j$ doesn't depend on the boundary  $\alpha_i$ whereas $\hat{W}_j$ does. The $\hat{W}_j$ also evolve degrees of freedom inside the $i$th measurement's future light cone. Consequently, the $\hat{W}_j$ commute with the $\hat{V}_j$, allowing us to simplify the expectation value in Eq. (\ref{eq:generalPB}) to
\begin{equation*}
\left\langle \hat{U}_{1}^{\dagger}\hat{P}_{1}\left(\alpha_{1}\right)...\hat{U}_{i}^{\dagger}\hat{P}_{i}\left(\alpha_{i}\right)\hat{V}_{i+1}^{\dagger}...\hat{P}_{n}\left(\alpha_{n}\right)\hat{V}_{n+1}^{\dagger}\hat{W}^{\dagger}\hat{P}_{\beta}\hat{W}\hat{V}_{n+1}\hat{P}_{n}\left(\alpha_{n}\right)...\hat{V}_{i+1}\hat{P}_{i}\left(\alpha_{i}\right)\hat{U}_{i}...\hat{P}_{1}\left(\alpha_{1}\right)\hat{U}_{1}\right\rangle ,
\end{equation*}
where the expectation value is taken over $\ket{\rm ini}$ and $\hat{W}\equiv\hat{W}_{n+1}...\hat{W}_{i+1}$. Since $\hat{W}^\dagger$ commutes with $\hat{P}_\beta$, it can be moved to the right of it where it will act on $\hat{W}$ and result in the identity matrix. Similarly, $\hat{P}_i(\alpha_i)$ can be moved to the right of $\hat{P}_\beta$ and we obtain
\begin{equation*}
p\left(\beta|\tilde{\Omega}\right)=\sum_{\alpha_{1},...,\alpha_{n}}\left\langle \hat{U}_{1}^{\dagger}\hat{P}_{1}\left(\alpha_{1}\right)...\hat{U}_{i}^{\dagger}\hat{V}_{i+1}^{\dagger}...\hat{P}_{n}\left(\alpha_{n}\right)\hat{V}_{n+1}^{\dagger}\hat{P}_{\beta}\hat{V}_{n+1}\hat{P}_{n}\left(\alpha_{n}\right)...\hat{V}_{i+1}\hat{P}_{i}\left(\alpha_{i}\right)\hat{U}_{i}...\hat{P}_{1}\left(\alpha_{1}\right)\hat{U}_{1}\right\rangle 
\end{equation*}
The sum over $\alpha_i$ can then be directly applied on $\hat{P}_i(\alpha_i)$ allowing us to eliminate it. As a result, $p\left(\beta|\tilde{\Omega}\right)$ is independent of basis chosen during the $i$th measurement.

The above argument can be applied sequentially and in reverse chronological order to eliminate $p\left(\beta|\tilde{\Omega}\right)$'s dependence on all bases associated with measurements outside Bob's measurement's past light cone. Although we conducted our analysis in a particular reference frame, and so assuming a particular ordering of events, our arguments could be applied to any other reference frame (modulo a relabeling of spacetime points). We would always arrive to the same conclusion: the causal-conditional prescription does not violate the no-signaling condition.

\section{Conclusions}

We have shown that modifying linear quantum mechanics is not as simple as adding terms in the Hamiltonian that depend on the wavefunction. One must also make a choice on how to interpret measurements and the evolution of the wavefunction. By breaking linearity, different formulations of quantum mechanics, such as the Everett and Copenhagen interpretations, no longer make equivalent predictions. 

By introducing the notion of a time-evolution operator that depends on the specified boundary conditions for the quantum state of the system that is being time-evolved, we were able to explore the range of possible prescriptions for assigning probabilities to measurement outcomes in NLQM. 
For a certain class of non-linear theories, we showed that two reasonable prescriptions do not violate the no-signaling condition.
The first is the Everett interpretation, and the second, which we named \emph{causal-conditional}, states that a measurement event at a particular spacetime point $\bm X$ updates the boundary state associated with the time  evolution operator of quantum degrees of freedom lying in the future light cone of $\bm X$. 
The predictions of causal-conditional are mathematically equivalent to standard quantum mechanics with causal feedback.
A measurement applies a feedback force (the details of which are determined by the non-linear theory of interest) on degrees of freedom lying in the future light cone of that measurement event.

\section*{Ackowledgments}
We thank Craig Savage, Sabina Scully, Tian Wang, Maneeli Derakhshani, Yiqiu Ma, Haixing Miao, Sean Carroll and Ashmeet Singh for useful discussions. We also thank Antoine Tilloy for pointing out important issues with how we explained certain concepts. 
This research is supported by NSF grants PHY-1404569 and PHY-1506453, as well as the Institute for Quantum Information and Matter, a Physics Frontier Center.

\section*{Appendix: The statistics of measurement outcomes in different reference frames}
We first review why sQM predicts, in all reference frames, identical statistics for measurement results. We then show that this is no longer true in NLQM if we adopt an interpretation where the wavefunction instantaneously collapses across all of space. Finally, we discuss why the causal-conditional prescription predicts, in different reference frames, identical statistics for measurement outcomes.

Consider the multiple measurements configuration shown in Fig. (\ref{fig:setup_measurements}) where Alice and then Bob measure their respective particles, as viewed in some reference frame that what we'll refer to as the \textit{lab frame}. In this frame, the probability of Alice measuring $\alpha$ and Bob measuring $\beta$, $p\left(\alpha,\beta\right|A_a, B_b)$ is given by Eqs. (\ref{eq:expressionProb}, \ref{eq:condStateComplicated}).
To conveniently transform $p\left(\alpha,\beta\right|A_a, B_b)$ from one reference
frame to another, we will express $p\left(\alpha,\beta\right|A_a, B_b)$ in a Heisenberg picture. Define 
\begin{equation}
\hat{P}_{\alpha}\left(\bm{x}_{A},t_{1}\right)\equiv\hat{U}^{\dagger}\left(t_{1},t_{0}\right)\hat{P}_{\alpha}\hat{U}\left(t_{1},t_{0}\right),\quad\hat{P}_{\beta}\left(\bm{x}_{B},t_{2}\right)\equiv\hat{U}^{\dagger}\left(t_{2},t_{0}\right)\hat{P}_{\beta}\hat{U}\left(t_{2},t_{0}\right),\label{eq:HeisProj}
\end{equation}
where we've explicitly denoted the location of Alice's measurement
at $\bm{x}_{A}$ and of Bob's at $\bm{x}_{B}$.
Since $\hat{P}_{\alpha}\left(\bm{x}_{A},t_{1}\right)$ and $\hat{P}_{\beta}\left(\bm{x}_{B},t_{2}\right)$
commute, we can rewrite $p\left(\alpha,\beta|A_a, B_b\right)$ to 
\begin{equation}
p\left(\alpha,\beta\right|A_a, B_b)=\braOket{\Psi_{{\rm ini}}}{\hat{P}_{\beta}\left(\bm{x}_{B},t_{2}\right)\hat{P}_{\alpha}\left(\bm{x}_{A},t_{1}\right)}{\Psi_{{\rm ini}}}.
\end{equation}

Consider now a Lorentz-transformation $\Lambda$ from the lab frame
to any other frame. On the Hilbert space of Alice and Bob's particles',
$\Lambda$ is realized by an operator $\hat{V}\left(\Lambda\right)$.
For instance, $\hat{V}\left(\Lambda\right)$ transforms a momentum
eigenstate of a spinless particles to $\hat{V}\left(\Lambda\right)\left|\bm{k}\right\rangle =\left|\Lambda\bm{k}'\right\rangle ,$
where $\left|\bm{k}\right\rangle $ is covariantly normalized to $\left\langle \bm{k}|\bm{k}'\right\rangle =\left\langle \Lambda\bm{k}|\Lambda\bm{k}'\right\rangle $ \cite{duncan2012the}. 
We re-express $p\left(\alpha,\beta|A_a, B_b\right)$ in terms of wavefunctions and projection operators viewed in a different frame 
\begin{eqnarray}
p\left(\alpha,\beta|A_a, B_b\right) & = & \braOket{\Lambda\Psi_{{\rm ini}}}{\hat{V}\left(\Lambda\right)\hat{P}_{\beta}\left(\bm{x}_{B},t_{2}\right)\hat{V}^{\dagger}\left(\Lambda\right)\hat{V}\left(\Lambda\right)\hat{P}_{\alpha}\left(\bm{x}_{A},t_{1}\right)\hat{V}^{\dagger}\left(\Lambda\right)}{\Lambda\Psi_{{\rm ini}}}\\
 & = & \braOket{\Lambda\Psi_{{\rm ini}}}{\hat{P}_{\beta}\left(\Lambda_{\,\nu}^{\mu}x_{B}^{\nu}\right)\hat{P}_{\alpha}\left(\Lambda_{\,\nu}^{\mu}x_{A}^{\nu}\right)}{\Lambda\Psi_{{\rm ini}}},\label{eq:pABreexpressed}
\end{eqnarray}
where $\left|\Lambda\Psi\right\rangle \equiv\hat{V}\left(\Lambda\right)\left|\Psi\right\rangle$ for any $\ket{\Psi}$, $x_{A}^{\nu}$ is the 4-vector $\left(\bm{x}_{A},t_{1}\right)$ and $x_{B}^{\nu}$ is $\left(\bm{x}_{B},t_{2}\right)$. If we assume that the measured results do not change under Lorentz transformations (\emph{e.g.} photodetector clicks\footnote{For a Klein-Gordon field $\hat{\phi}$, the measured observable would be 
\begin{equation}
\int_{V}\hat{j}^{\nu}.d\Sigma_{\nu},\quad\hat{j}^{\nu}\left(x^{\mu}\right)=i\left[\partial^{\nu}\hat{\phi}_{-}\left(x^{\mu}\right)\right]\hat{\phi}_{+}\left(x^{\mu}\right)+h.c,
\end{equation}
where $V$ is the spacetime volume occupied by the photodetector during
a single measurement run, and $\hat{\phi}_{+}$ and $\hat{\phi}_{-}$
are the positive and negative frequency components of $\hat{\phi}$,
respectively \cite{belinda}.}), then Eq. (\ref{eq:pABreexpressed}) is just the probability of
measuring $\alpha$ and $\beta$ in a different reference frame. Therefore,
in sQM, the statistics of measurement outcomes are the same in all
reference frames.

The extension of $p\left(\alpha,\beta\right|A_a, B_b)=\left\langle \Psi_{c|\alpha,\beta}|\Psi_{c|\alpha,\beta}\right\rangle $, as calculated by an observer in the lab frame, to NLQM coupled with an interpretation of QM with wavefunction collapse
is
\begin{equation}
\left|\Psi_{c|\alpha,\beta}^{{\rm collapse}}\right\rangle =\hat{P}_{\beta}\hat{U}_{\Phi_{\alpha}\left(t_{1}\right)}\left(t_{2},t_{1}\right)\hat{P}_{\alpha}\hat{U}_{\Psi_{{\rm ini}}\left(t_{0}\right)}\left(t_{1},t_{0}\right)\left|\Psi_{{\rm ini}}\right\rangle ,
\end{equation}
where $\hat{U}_{\Psi_{{\rm ini}}\left(t_{0}\right)}\left(t_{1},t_{0}\right)$
is the time-evolution operator associated with the boundary condition
$\left|\Psi\left(t_{0}\right)\right\rangle =\left|\Psi_{{\rm ini}}\right\rangle $,
and $\hat{U}_{\Phi_{\alpha}\left(t_{1}\right)}\left(t_{2},t_{1}\right)$
is associated with the condition $\left|\Psi\left(t_{1}\right)\right\rangle =\left|\Phi_{\alpha}\right\rangle $
where 
\begin{equation}
\left|\Phi_{\alpha}\right\rangle =\hat{P}_{\alpha}\hat{U}_{\Psi_{{\rm ini}}\left(t_{0}\right)}\left(t_{1},t_{0}\right)\left|\Psi_{{\rm ini}}\right\rangle .
\end{equation}
The extension of the Heisenberg picture projection operators in Eq.
(\ref{eq:HeisProj}) to NLQM are
\begin{eqnarray}
\hat{P}_{\alpha}^{{\rm collapse}}\left(\bm{x}_{A},t_{1}\right) & \equiv & \hat{U}_{\Psi_{{\rm ini}}\left(t_{0}\right)}^{\dagger}\left(t_{1},t_{0}\right)\hat{P}_{\alpha}\hat{U}_{\Psi_{{\rm ini}}\left(t_{0}\right)}\left(t_{1},t_{0}\right),\quad\hat{P}_{\beta}^{{\rm collapse}}\left(\bm{x}_{B},t_{2}\right)\equiv\hat{U}_{2}^{\dagger}\hat{P}_{\beta}\hat{U}_{2},\\
\hat{U}_{2} & \equiv & \hat{U}_{\Phi_{\alpha}\left(t_{1}\right)}\left(t_{2},t_{1}\right)\hat{U}_{\Psi_{{\rm ini}}\left(t_{0}\right)}\left(t_{1},t_{0}\right).\label{eq:U2NLdef}
\end{eqnarray}
Consequently, the extension of Eq. (\ref{eq:pABreexpressed}) to NLQM
is 
\begin{equation}
p^{{\rm collapse}}\left(\alpha,\beta\right|A_a, B_b)=\braOket{\Lambda\Psi_{{\rm ini}}}{\hat{P}_{\beta}^{{\rm collapse}}\left(\Lambda_{\,\nu}^{\mu}x_{B}^{\nu}\right)\hat{P}_{\alpha}^{{\rm collapse}}\left(\Lambda_{\,\nu}^{\mu}x_{A}^{\nu}\right)}{\Lambda\Psi_{{\rm ini}}}.
\end{equation}

Let's consider a Lorentz transformation $\tilde{\Lambda}$ that takes
the lab frame to one where Bob measures his particles before Alice
measures hers. An observer in that frame would calculate that the
probability that Alice measures $\alpha$ and Bob measures $\beta$
is 
\begin{eqnarray*}
\tilde{p}^{{\rm collapse}}\left(\alpha,\beta\right|A_a, B_b) & = & \braOket{\tilde{\Lambda}\Psi_{{\rm ini}}}{\tilde{P}_{\alpha}^{{\rm collapse}}\left(\tilde{x}_{A}\right)\tilde{P}_{\beta}^{{\rm collapse}}\left(\tilde{x}_{B}\right)}{\tilde{\Lambda}\Psi_{{\rm ini}}},\quad\tilde{x}_{A,B}\equiv\tilde{\Lambda}_{\,\nu}^{\mu} x_{A,B}^{\nu}, \\
\tilde{P}_{\beta}^{{\rm collapse}}\left(\tilde{x}_{B}\right) & = & \hat{U}_{\tilde{\Lambda}\Psi_{{\rm ini}}\left(t_{0}\right)}^{\dagger}\left(\tilde{x}_{B}^{0},\tilde{t}_{0}\right)\hat{P}_{\beta}\hat{U}_{\tilde{\Lambda}\Psi_{{\rm ini}}\left(t_{0}\right)}\left(\tilde{x}_{B}^{0},\tilde{t}_{0}\right),\quad\tilde{P}_{\alpha}^{{\rm collapse}}\left(\tilde{x}_{A}\right)=\tilde{U}_{2}^{\dagger}\hat{P}_{\alpha}\tilde{U}_{2},\\
\tilde{U}_{2} & \equiv & \hat{U}_{\tilde{\Phi}_{\beta}\left(\tilde{x}_{B}^{0}\right)}\left(\tilde{x}_{A}^{0},\tilde{x}_{B}^{0}\right)\hat{U}_{\tilde{\Lambda}\Psi_{{\rm ini}}\left(t_{0}\right)}\left(\tilde{x}_{B}^{0},\tilde{t}_{0}\right), \\ 
\left|\tilde{\Phi}_{\beta}\right\rangle & = &\hat{P}_{\beta}\hat{U}_{\tilde{\Lambda}\Psi_{{\rm ini}}\left(t_{0}\right)}\left(\tilde{x}_{B}^{0},\tilde{t}_{0}\right)\left|\tilde{\Lambda}\Psi_{{\rm ini}}\right\rangle ,
\end{eqnarray*}
where $\tilde{t}_{0}$ is when the experiment began in this new frame.
Although $p^{{\rm collapse}}\left(\alpha,\beta|A_a, B_b\right)$ can be re-written
to 
\begin{equation}
p^{{\rm collapse}}\left(\alpha,\beta|A_a, B_b\right)=\braOket{\tilde{\Lambda}\Psi_{{\rm ini}}}{\hat{P}_{\alpha}^{{\rm collapse}}\left(\tilde{x}_{A}\right)\hat{P}_{\beta}^{{\rm collapse}}\left(\tilde{x}_{B}\right)}{\tilde{\Lambda}\Psi_{{\rm ini}}},
\end{equation}
it isn't in general equal to $\tilde{p}^{{\rm collapse}}\left(\alpha,\beta|A_a, B_b\right)$
because $\hat{P}_{\beta}^{{\rm collapse}}\left(\tilde{x}_{B}\right)$
depends on $\Phi_{\alpha}$ while $\tilde{P}_{\beta}^{{\rm collapse}}\left(\tilde{x}_{B}\right)$
doesn't. Similarly, $\tilde{P}_{\alpha}^{{\rm collapse}}\left(\tilde{x}_{A}\right)$
depends on $\tilde{\Phi}_{\beta}$ while $\hat{P}_{\alpha}^{{\rm collapse}}\left(\tilde{x}_{A}\right)$
doesn't. In other words, $\hat{V}\left(\tilde{\Lambda}\right)$ doesn't
connect $\Phi_{\alpha},$ $\Phi_{\beta}$ and $\Psi_{{\rm ini}}$
to each other.

The fact that $p^{{\rm collapse}}\left(\alpha,\beta|A_a, B_b\right)$ isn't
the same in all reference frame isn't surprising. and can be understood
heuristically when we view the non-linearity as a feedback force that
changes acausally after the wavefunction collapses. Consider the
following non-linear interaction energy density 
\begin{equation}
\hat{V}_{{\rm NL}}\left(\bm{x}\right)=\braOket{\Psi\left(t\right)}{\hat{O}}{\Psi\left(t\right)}\hat{M}\left(\bm{x}\right),
\end{equation}
where $\hat{O}$ is Lorentz-invariant ($\hat{V}^{\dagger}\left(\Lambda\right)\hat{O}\hat{V}\left(\Lambda\right)=\hat{O}$
for any transformation $\Lambda$), and we assume that $\hat{M}\left(x\right)$
transforms as $\hat{M}\left(\Lambda x\right)$ under $\Lambda$ (so
as to maintain the the requirement that the total interaction Hamiltonian
density $\hat{H}_{{\rm int}}\left(x\right)$ transforms as $\hat{H}_{{\rm int}}\left(\Lambda x\right)$
- see sec. 5.5 of \cite{duncan2012the}). We can then view $F\left(t\right)\equiv\braOket{\Psi\left(t\right)}{\hat{O}}{\Psi\left(t\right)}$
as a classical feedback force on $\hat{M}$.

When a measurement occurs, $\Psi(t)$  instantaneously changes, and
so $F\left(t\right)$ acting on $\hat{M}\left(\bm{x}\right)$ for all $\bm{x}\in\mathbb{R}^{3}$
changes instantaneously too. The
problem is that the spatial surface of time simultaneity is the not
the same in all reference frames. In the case of multiple spacelike-separated
measurement events, we'd get that $F\left(t\right)$ changes differently
in different frames depending on the ordering of the measurement events
in that frame. On the other hand, for the causal-conditional prescription,
$F\left(t\right)$ would change causally after any measurement, and
so there are no issues.

\section*{References}

\bibliographystyle{plain}
\bibliography{references}

\end{document}